\documentclass[british,english]{article}
\usepackage[T1]{fontenc}
\usepackage[latin9]{inputenc}
\usepackage{geometry}
\usepackage{babel}
\usepackage{amsmath}
\usepackage{amssymb}
\usepackage{graphicx}
\usepackage[authoryear]{natbib}
\usepackage[unicode=true,pdfusetitle,
 bookmarks=true,bookmarksnumbered=false,bookmarksopen=false,
 breaklinks=false,pdfborder={0 0 1},backref=false,colorlinks=false]
 {hyperref}

\makeatletter

\usepackage{hyperref}
\usepackage[linesnumbered,ruled]{algorithm2e}
\usepackage{authblk}

\@ifundefined{showcaptionsetup}{}{%
 \PassOptionsToPackage{caption=false}{subfig}}
\usepackage{subfig}
\makeatother

\usepackage{babel}
\begin{document}
\title{Sequential Monte Carlo with active subspaces}
\author[1]{Leonardo Ripoli}
\author[2,3]{Richard G. Everitt}
\affil[1]{Department of Mathematics and Statistics, University of Reading}
\affil[2]{Department of Statistics, University of Warwick}
\affil[3]{The Zeeman Institute for Systems Biology \& Infectious Disease Epidemiology Research, University of Warwick}
\affil[2]{Email: \href{mailto:richard.everitt@warwick.ac.uk}{richard.everitt@warwick.ac.uk}}
\maketitle
\begin{abstract}
Monte Carlo methods, such as Markov chain Monte Carlo (MCMC), remain
the most regularly-used approach for implementing Bayesian inference.
However, the computational cost of these approaches usually scales
worse than linearly with the dimension of the parameter space, limiting
their use for models with a large number of parameters. However, it
is not uncommon for models to have identifiability problems. In this
case, the likelihood is not informative about some subspaces of the
parameter space, and hence the model effectively has a dimension that
is lower than the number of parameters. \citet{constantine_accelerating_2016,schuster_exact_2017}
introduced the concept of directly using a Metropolis-Hastings (MH)
MCMC on the subspaces of the parameter space that are informed by
the likelihood as a means to reduce the dimension of the parameter
space that needs to be explored. The same paper introduces an approach
for identifying such a subspace in the case where it is a linear transformation
of the parameter space: this subspace is known as the \emph{active
subspace}. This paper introduces sequential Monte Carlo (SMC) methods
that make use of an active subspace. As well as an SMC counterpart
to MH approach of \citet{schuster_exact_2017}, we introduce an approach
to learn the active subspace adaptively, and an SMC$^{2}$ approach
that is more robust to the linearity assumptions made when using active
subspaces.
\end{abstract}

\section{Introduction}

\subsection{Motivation}

This paper considers the Bayesian estimation of the random vector
parameter $\theta$ (with range $\mathbb{R}^{d}$) from data $y\in\mathcal{Y}$.
We let $p$ denote the prior distribution on $\theta$, and $f$ denote
the model for $y$ given $\theta$. Let $\pi$ denote the posterior
distribution on $\theta$, with $\tilde{\pi}$ denoting the unnormalised
version of this distribution arising from multiplying the prior and
likelihood, i.e. $\tilde{\pi}\left(\theta\mid y\right)=p\left(\theta\right)f\left(y\mid\theta\right)$
with $\pi\left(\theta\mid y\right)=\tilde{\pi}\left(\theta\mid y\right)/p\left(y\right)$.
From hereon we denote the posterior as $\pi\left(\theta\right):=\pi\left(\theta\mid y\right)$
likelihood as $l\left(\theta\right):=f\left(y\mid\theta\right)$.
In this paper we require that $l$ is differentiable in $\theta$.
We let $Z$ denote the normalising constant of $\tilde{\pi}$.

It is not uncommon for models to have identifiability problems \citep{constantine_accelerating_2016}.
In this case, the likelihood is not informative about some subspaces
of the parameter space, and hence the model has an `intrinsic' dimension
that is lower than the number of parameters. \citet{constantine_accelerating_2016}
introduces the idea of identifying the subspace of the parameter space
that is informed by the likelihood (the `active' subspace (AS)),
and using MCMC on this space only. The advantage of this approach,
when it is possible, is that the dimension of the AS is smaller than
the dimension of the whole space, therefore we expect that our MCMC
will not need to be run for as many iterations. This paper makes three
contributions:
\begin{itemize}
\item We introduce the \emph{active subspace sequential Monte Carlo} (AS-SMC)
algorithm, which makes use of active subspaces in sequential Monte
Carlo (SMC).
\item In most previous work the identification of an AS must be performed
in advance of running an inference algorithm. A typical approach would
be to simulate points from the prior, and to use these to infer an
AS. However, as we outline in the next section, were they available,
one would use points from the posterior to infer the AS. This paper
introduces the \emph{adaptive AS-SMC} approach to allow the adaptation
of the AS as the algorithm runs, using the points simulated at each
step of the algorithm.
\item We further develop the AS-SMC algorithm, introducing a nested SMC
approach we call AS-SMC$^{2}$.
\end{itemize}

\subsection{Active subspace Metropolis-Hastings}

\subsubsection{Active subspaces}

We follow the approach of \citet{constantine_accelerating_2016} and
\citet{schuster_exact_2017}, which we now describe. This approach
considers only subspaces that are linear projections of the parameter
space, and this is also the approach used in this paper. We denote
the \emph{active variables} by $a$ (in $\mathbb{R}^{d_{a}})$ and
\emph{inactive variables} by $i$ (in $\mathbb{R}^{d_{i}})$. Our
aim is to find a reparameterisation of $\theta$ of the form
\begin{equation}
\theta=Aa+Ii,
\end{equation}
where $A\in\mathbb{R}^{d\times d_{a}}$ and $I\in\mathbb{R}^{d\times d_{i}}$
are such that $\left[A,I\right]$ is an orthonormal basis of $\mathbb{R}^{d}$.
We wish to choose the reparameterisation such that performing MCMC
on the marginal distribution of $a$ is sufficient for simulation
from the posterior $\pi$. \citet{constantine_accelerating_2016}
also describes an approach for finding such a reparameterisation.
The key idea is to find an estimate of the eigendecomposition of the
matrix
\begin{equation}
\int_{\theta}\nabla\log l\left(\theta\right)\nabla\log l\left(\theta\right)^{T}\phi\left(\theta\right)d\theta,\label{eq:as}
\end{equation}
for some distribution $\phi$ (the choice of which will be discussed
in section \ref{subsec:Prior-and-posterior}), using an eigendecomposition
of the Monte Carlo approximation
\begin{equation}
\frac{1}{N}\sum_{m=1}^{N}\nabla\log l\left(\theta^{m}\right)\nabla\log l\left(\theta^{m}\right)^{T}\label{eq:est_as}
\end{equation}
for $\theta^{i}\sim\phi$ for $m=1:N$. This is equivalent to performing
a uncentred and unscaled principal component analysis (PCA) to find
the directions in which the score function $\nabla\log l\left(\theta\right)$
is most variable under $\phi$: the eigenvectors of this decomposition
with the largest eigenvalues will indicate the directions in which
there is most variation in $\nabla\log l\left(\theta\right)$. The
directions in which there is little variation in $\nabla\log l\left(\theta\right)$
- when the likelihood is flat - are those that we may designate as
inactive variables. \citet{constantine_accelerating_2016,schuster_exact_2017}
propose to identify a spectral gap in the eigendecomposition to choose
which directions are designated as active and inactive variables:
identifying a group of dominant eigenvalues that are well separated
from the remainder, and choosing the corresponding directions to give
the active variables. This procedure is similar to the approach used
in PCA for dimensionality reduction, where the directions that explain
a large proportion (e.g. 90\%) of the variance are selected.

\subsubsection{Active subspace Metropolis-Hastings\label{subsec:Active-subspace-Metropolis-Hasti}}

To see how to use active subspaces within MCMC, we first write the
posterior on the new parameterisation.
\begin{eqnarray}
\pi_{a,i}\left(a,i\right) & \propto & p_{a,i}\left(a,i\right)l\left(Aa+Ii\right),\label{eq:as_posterior}
\end{eqnarray}
where $p_{a,i}\left(a,i\right)=p\left(Aa+Ii\right)$. Since $B_{a}$
and $B_{i}$ are orthonormal there is no Jacobian associated with
this change in parameterisation. Throughout, as in \citet{constantine_accelerating_2016},
we assume that we have available the distributions $p_{a}$ and $p_{i}$
arising from the factorisation $p_{a,i}\left(a,i\right)=p_{a}\left(a\right)p_{i}\left(i\mid a\right)$:
specifically, that we can evaluate $p_{a}$ and $p_{i}$ pointwise,
and that we can simulate from $p_{i}\left(\cdot\mid a\right)$. \citet{constantine_accelerating_2016}
describes how to construct an approximate MCMC algorithm (in the style
of \citet{Alquier2016}) on the active variables through using a numerical
estimate of the (unnormalised) marginal distribution
\begin{eqnarray*}
\tilde{\pi}_{a}\left(a\right) & = & \int_{i}p\left(Aa+Ii\right)l\left(Aa+Ii\right)di\\
 & = & p_{a}\left(a\right)\int_{i}p_{i}\left(i\mid a\right)l\left(Aa+Ii\right)di
\end{eqnarray*}
at every iteration of the algorithm. The second term is the marginal
likelihood
\begin{equation}
l_{a}\left(a\right):=\int_{i}p_{i}\left(i\mid a\right)l\left(Aa+Ii\right)di.\label{eq:llhd}
\end{equation}
\citet{schuster_exact_2017} made the observation that this approximate
method can be made exact by formulating it as a pseudo-marginal algorithm
\citep{Beaumont2003,Andrieu2009}, using the unbiased importance sampling
(IS) estimator
\begin{equation}
\bar{l}_{a}\left(a\right)=\frac{1}{N_{i}}\sum_{n=1}^{N_{i}}\frac{p_{i}\left(i^{n}\mid a\right)l\left(Aa+Ii^{n}\right)}{q_{i}\left(i^{n}\mid a\right)},\label{eq:llhd_estimator}
\end{equation}
where $i^{n}\sim q_{i}\left(\cdot\mid a\right)$ for $n=1:N_{i}$,
at each step of the MCMC. Algorithm \ref{alg:AS-MH} gives the active
subspace MH (AS-MH) algorithm from \citet{schuster_exact_2017}).

\begin{algorithm}
\caption{Active subspace Metropolis-Hastings}\label{alg:AS-MH}

Initialise $a^0$;\

\For {$n=1:N_i$}
{
	$i^{n,0} \sim q_{i} \left(\cdot \mid a^0 \right)$;\

	\[
	\tilde{w}^{n,0} = \frac{p_{i}\left(i^{n,0} \mid a^{0} \right)l\left( Aa^0 + Ii^{n,0}\right)}{q_{i}\left(i^{n,0}\mid a^0\right)};
	\]\
}

$u^{0} \sim\mathcal{M}\left( \left( w^{1,0}, ..., w^{N_i,0} \right) \right),$ where for $n=1:N_i$
\[
w^{n,0} = \frac{\tilde{w}^{n,0}}{\sum_{p=1}^{N_{i}} \tilde{w}^{p,0}};\
\]

Let $\bar{l}^0_{a}=\frac{1}{N_{i}}\sum_{n=1}^{N_{i}} \tilde{w}^{n,0}$;\

\For {$m=1:N_a$}
{
	$ a^{*m} \sim q_a\left(\cdot\mid a^{m-1} \right)$;\

	\For {$n=1:N_i$}
	{

		$i^{*n,m} \sim q_{i} \left(\cdot \mid a^{*m} \right);$\

		\[
		\tilde{w}^{*n,m} = \frac{p_{i}\left(i^{*n,m} \mid a^{*m} \right) l\left( A a^{*m} + I i^{*n,m}\right)}{q_{i}\left( i^{*n,m}\mid a^{*m} \right)};
		\]\
	}

	$u^{*m} \sim\mathcal{M}\left( \left( w^{*1,m}, ..., w^{*N_i,m} \right) \right),$ where for $n=1:N_i$
	\[
	w^{*n,m} = \frac{\tilde{w}^{*n,m}}{\sum_{p=1}^{N_{i}} \tilde{w}^{*p,m}};\
	\]

	Let $\bar{l}_{a}\left(a^{*m} \right)=\frac{1}{N_{i}}\sum_{n=1}^{N_{i}} \tilde{w}^{*n,m} $;\

    Set $\left(a^m, \left\{ i^{n,m}, w^{n,m} \right\}_{n=1}^{N_i}, u^m, \bar{l}^m_{a} \right) = \left(a^{*m},\left\{ i^{*n,m}, w^{*n,m} \right\}_{n=1}^{N_i}, u^{*m}, \bar{l}_{a}\left(a^{*m} \right) \right)$ with probability
	\[
	\alpha_a^m = 1\wedge\frac{p_{a}\left(a^{*m} \right) \bar{l}_{a}\left(a^{*m}\right)}{p_{a}\left(a^{m-1} \right) \bar{l}^{m-1}_{a}}\frac{q_a\left(a^{m-1} \mid a^{*m}\right)}{q_a\left(a^{*m}\mid a^{m-1} \right)};
	\]\
	
	Else let $\left(a^m, \left\{ i^{n,m}, w^{n,m} \right\}_{n=1}^{N_i}, u^m, \bar{l}^m_{a} \right) = \left(a^{m-1},\left\{ i^{n,m-1}, w^{n,m-1} \right\}_{n=1}^{N_i}, u^{m-1}, \bar{l}^{m-1}_{a}\right)$;\

}
\end{algorithm}

\subsection{Efficiency of AS-MH\label{subsec:Prior-and-posterior}}

The statistical efficiency of AS-MH depends on the choice of $q_{a},q_{i}$
and the AS defined by $A$ and $I$. For a pseudo-marginal approach
to offer better performance than an MH algorithm that directly proposes
a move on the vector $\theta$, we require that the estimator in equation
(\ref{eq:llhd_estimator}) has a low variance, so that the pseudo-marginal
resembles the marginal MH, where the integral in equation (\ref{eq:llhd})
is available exactly. The optimal choice of $q_{i}\left(i\mid a\right)$,
in terms of minimising the IS variance, would be for it to be proportional
to $p_{i}\left(i\mid a\right)l\left(Aa+Ii\right)$. We now describe
how the use of an AS allows us to use a proposal close to this optimal
choice.

The AS is chosen so that, in an ideal case, the inactive variables
are not at all influenced by the likelihood. In this case, for any
$a$, $l\left(Aa+Ii\right)\propto1$ as $\theta_{i}$ varies, thus
the optimal choice of $q_{i}\left(i\mid a\right)$ is the prior $p_{i}\left(i\mid a\right)$.
Since this distribution is available to us, in this ideal case we
can easily implement an efficient pseudo-marginal method. We can see
the AS as offering a criterion through which we may determine when
the pseudo-marginal approach, with the prior $p_{i}$ as the proposal,
might be an effective sampler.

Outside of the ideal case, the likelihood may only be approximately
flat as $i$ varies, thus $p_{i}$ will not be optimal, with the distance
from optimality depending on $a$. We now consider two ways in which
the AS-MH algorithm may be improved, each of which motivates a method
in the paper.

\subsubsection{Prior and posterior active subspaces}

For the AS-MH algorithm to be effective, we require that $p_{i}$
is close to optimal for the values of $a$ that we might visit when
running the algorithm: i.e. those that are the region of the posterior
$\pi_{a}$. Recall equation (\ref{eq:as}), used to find the AS: here
we aimed to find the directions in which the score function $\nabla\log l\left(\theta\right)$
is most variable under some distribution $\phi$. We now see the effect
that the choice of $\phi$ (and hence $B_{a}$ and $B_{i}$) might
have on the efficiency of AS-MH: if $\phi$ is close to the posterior
$\pi$, $p_{i}$ is likely to be a good proposal for the inactive
variables conditional on the $a$ we are likely to visit. If $\phi$
is far from $\pi$, then although the inactive variables will be chosen
to be those directions in which $\nabla\log l\left(\theta\right)$
is least variable over $\phi$, the directions may not correspond
to the directions in which $\nabla\log l\left(\theta\right)$ is least
variable over $\pi$. This suggests $\phi=\pi$ as an appropriate
choice for determining an AS for used in AS-MH: we refer to this as
the \emph{posterior active subspace}.

Recall now equation (\ref{eq:est_as}), the Monte Carlo approximation
used to estimate equation (\ref{eq:as}). In \citet{constantine_accelerating_2016},
$\phi$ is taken to be equal to the prior $p$: we call this the \emph{prior
active subspace}. Since it is usually possible to easily sample from
$p$, it is straightforward to implement equation (\ref{eq:est_as}).
However, the prior AS may not be close to the posterior AS. An alternative
to using equation (\ref{eq:est_as}) for estimating the posterior
AS would be to use an IS estimate with the prior as the proposal,
but this will only be accurate when the prior is close to the posterior,
in which case the prior and posterior active subspaces are unlikely
to differ significantly.

In this section we introduce a toy example that illustrates a case
when the prior and posterior active subspaces are very different.
Let the prior on $\theta$ be multivariate Gaussian $\theta\sim\mathcal{MVN}\left(\mathbf{0},5000\mathbf{I}\right)$
and the likelihood have the form
\[
l\left(\theta\right)\propto\prod_{j=1}^{d}\left[\exp\left(-\left(\frac{\theta_{j}}{\sigma_{j}}\right)^{2}\right)\frac{1}{1+\left(\frac{\theta_{j}}{\gamma_{j}}\right)^{2}}\right],
\]
where for the purposes of this definition only we have altered the
meaning of the subscript of $\theta$ to denote the different dimensions
of $\theta$. Each dimension of $\theta$ has an independent likelihood
term, each composed of a Gaussian and a Cauchy term. Its gradient
is
\[
\nabla\log l\left(\theta\right)=\left(\begin{array}{c}
-2\theta_{1}\left(\frac{1}{\sigma_{1}^{2}}+\frac{1}{\theta_{1}^{2}+\gamma_{1}^{2}}\right)\\
...\\
-2\theta_{d}\left(\frac{1}{\sigma_{d}^{2}}+\frac{1}{\theta_{d}^{2}+\gamma_{d}^{2}}\right)
\end{array}\right).
\]
As each $\theta_{j}$ gets further from 0, the Gaussian term ($1/\sigma_{1}^{2}$)
dominates, and so for $\theta_{j}$ far from zero this term will dominate
the gradient. However, when $\theta_{j}$ is near zero, with appropriately
set parameters, we will see the Cauchy term ($1/\left(\theta_{d}^{2}+\gamma_{d}^{2}\right)$)
dominate.

\begin{figure}
\subfloat[Prior and posterior in the toy example.\label{fig:Prior-and-posterior}]{\includegraphics[scale=0.3]{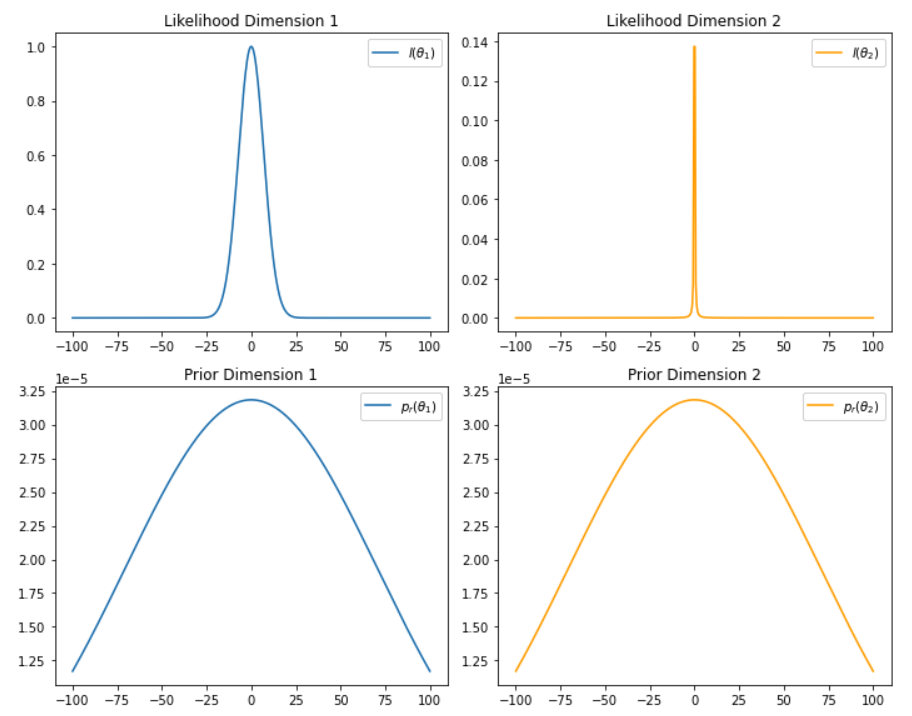}

}

\subfloat[\foreignlanguage{british}{PCA performed on the gradient of points simulated from the prior.\label{fig:PCA-performed-on}}]{\includegraphics[scale=0.25]{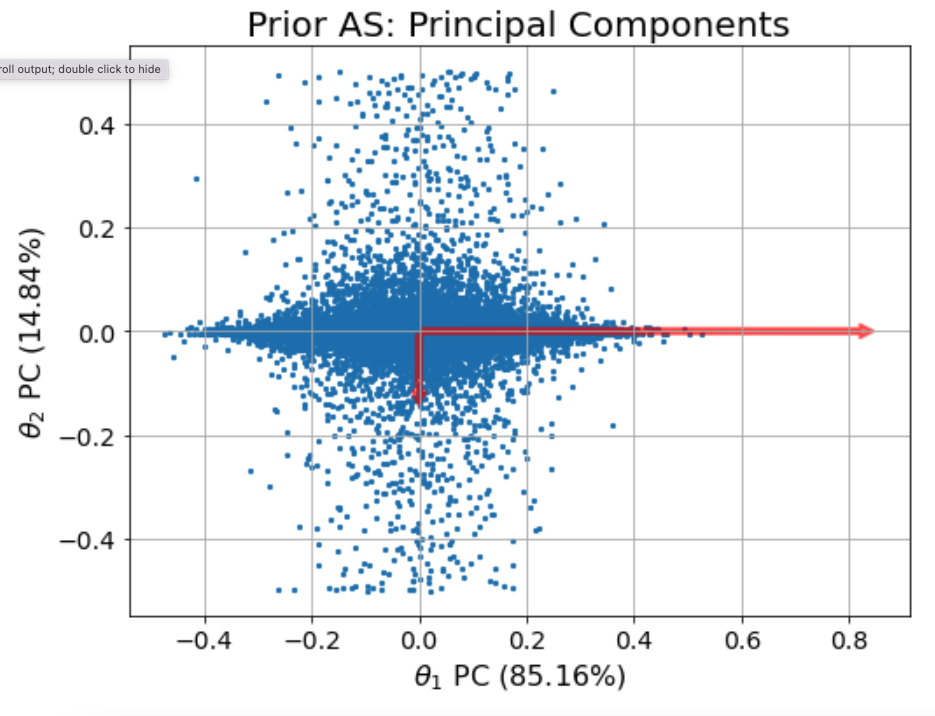}

}\subfloat[\foreignlanguage{british}{PCA performed on the gradient of points simulated from the posterior.\label{fig:PCA-performed-on-1}}]{\includegraphics[scale=0.25]{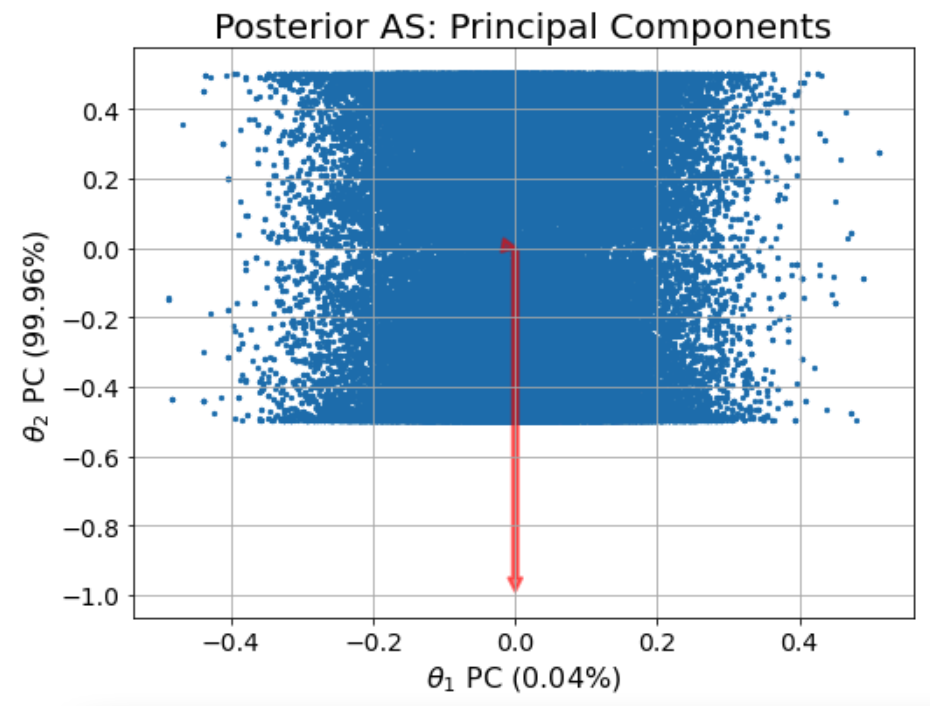}

}

\caption{\foreignlanguage{british}{An illustration of the toy example.\label{fig:An-illustration-of}}}
\end{figure}

We examine a two dimensional ($d=2$) example with $\sigma_{1}=10$,
$\gamma_{1}=10^{12}$, $\sigma_{2}=50$ and $\gamma_{2}=0.1$. Figure
\ref{fig:Prior-and-posterior} shows the prior and posterior for these
parameters, with figures \ref{fig:PCA-performed-on} and \ref{fig:PCA-performed-on-1}
showing uncentred and unscaled PCA performed on the gradient of points
simulated from the prior and posterior respectively. Across the range
of the prior, the gradient varies more in the first dimension, therefore
in the prior AS the second variable will be chosen to be inactive
through examining the spectral gap. However, in the posterior AS,
we see the first variable being chosen to be inactive.

\selectlanguage{english}%

\selectlanguage{british}%
We see from this example that the approach in \citet{constantine_accelerating_2016}
and \citet{schuster_exact_2017} of using the prior AS is not always
desirable. Whilst the example in this section is very artificial,
real applications can also exhibit significant differences between
the prior and posterior active subspaces (see \citet{parente_active_2020}
and \citet{zahm_certified_2022}, for example).

\subsubsection{Importance sampling for the inactive variables\label{subsec:Choosing-the-inactive}}

AS-MH uses the IS proposal $q_{i}$ to simulate the inactive variables
at every step of the MCMC. The prior $p_{i}$ is a good choice for
this in the case of `ideal' inactive variables. Outside of this
special case, it is less clear that using IS with this proposal should
be an efficient means for sampling these variables. active subspaces
will be of most use when the dimension of the inactive variables,
$d_{i}$, is large. Since IS requires an exponential number of points
in the dimension to obtain a fixed variance \citep{agapiou_importance_2017},
it is precisely the case of $d_{i}$ being large when IS will be most
inefficient. For $d_{i}$ large, the variance of the IS estimator
of the marginal likelihood is likely to be large, and hence AS-MH
is likely to be inefficient.

\subsection{Proposed approach and overview of paper}

In the paper we introduce AS-SMC as an SMC analogue of AS-MH, allowing
the use of active subspaces in SMC. This SMC sampler moves a population
of weighted importance points (`particles') through a sequence of
distributions, using an AS-MH move at each iteration. Typically this
approach would use the easy to estimate prior AS, which is useful
in a number of applications.

In some cases (such as in \citet{parente_active_2020} and \citet{zahm_certified_2022}),
an estimate of the posterior AS is required in order for AS-MH or
AS-SMC to be efficient. For this purpose we extend AS-SMC to, at each
iteration of the SMC, use an AS tailored to the current step of the
algorithm by estimating the AS using the current set of weighted particles
(section \ref{subsec:SMC-sampler-with-1}). If a useful AS is not
found at any given SMC iteration, a standard MH move will be used
instead of AS-MH. Adaptive sequential importance sampling approaches
such as this have been considered previously in \citet{parente_active_2020}
and \citet{zahm_certified_2022} These previous approaches adaptively
construct an AS in a similar way: our main contribution is to combine
such an approach with the exact pseudo-marginal framework introduced
by \citet{schuster_exact_2017}.

Separately we consider the case where the inactive variables have
some (small) influence on the likelihood, where IS is not an effective
method for estimating the marginal likelihood. For this situation
we introduce an SMC$^{2}$ algorithm, using an SMC algorithm to estimate
the marginal likelihood for each particle.

In section \ref{sec:SMC-with-active} we introduce the new methods,
followed by presenting empirical results in section \ref{sec:Empirical-results}
and conclusions in section \ref{sec:Conclusions}.

\section{SMC with active subspaces\label{sec:SMC-with-active}}

In section \ref{subsec:SMC-sampler-recap} we recap the fundamentals
of SMC samplers, before introducing SMC approaches using a fixed AS
in section \ref{subsec:SMC-sampler-with}, an SMC sampler using an
adaptive AS in section \ref{subsec:SMC-sampler-with-1}, followed
by the SMC$^{2}$ approach in section \ref{subsec:SMC-variant}.

\subsection{SMC sampler recap\label{subsec:SMC-sampler-recap}}

An SMC sampler \citep{DelMoral2006c} operates by using a sequence
of distributions $\pi_{0},...,\pi_{T}$ such that $\pi_{T}=\pi$,
$\pi_{0}$ is `easy' to simulate from, and where the sequence can
be seen as a path of distributions bridging between $\pi_{0}$ and
$\pi_{T}$. The algorithm is devised so as to guide a set of weighted
importance points (`particles') from $\pi_{0}$ to $\pi_{T}$. The
formal construction of the algorithm uses a sequence $\overline{\pi}_{0},...,\overline{\pi}_{T}$
of extended distributions, where for each $t$, $\overline{\pi}_{t}$
has $\pi_{t}$ as a marginal distribution. Specifically,
\begin{eqnarray}
\overline{\pi}_{t}\left(\theta_{0:t}\right) & = & \pi\left(\theta_{t}\right)\prod_{s=0}^{t-1}L_{s}\left(\theta_{s}\mid\theta_{s+1}\right),\label{eq:smc_extended_target}
\end{eqnarray}
where each $L_{s}$ is known as a `backwards' kernel. The SMC sampler
is a sequential importance sampling approach, which uses as a proposal
\begin{eqnarray}
\overline{q}_{t}\left(\theta_{0:t}\right) & = & q_{0}\left(\theta_{0}\right)\prod_{s=0}^{t}K_{s}\left(\theta_{s}\mid\theta_{s-1}\right)\label{eq:smc_extended_proposal}
\end{eqnarray}
at target $t$, where each $K_{s}$ is known as a `forwards' kernel.
The algorithm proceeds at iteration $t$ by moving each particle using
$K_{t}$, reweighting each particle using the ratio of equations (\ref{eq:smc_extended_target})
and (\ref{eq:smc_extended_proposal}), and resampling the weighted
population of particles so that particles with a small weight tend
not to remain in the population for the following iteration of the
algorithm. The marginal distribution of the weighted $\theta_{t}$
values of the particles can then be used as an approximation to $\pi_{t}$.

The implemention of this algorithm only requires the evaluation of
each $\pi_{t}$ up to a normalising constant; we let $\tilde{\pi}_{t}$
denote an unnormalised version of $\pi_{t}$. The weights calculated
at each step then contain information about the normalising constant
of each unnormalised distribution $\tilde{\pi}_{t}$, and can be used
to additionally calculate an estimate its normalising constant $Z_{t}$.
The choice of the sequence of distributions significantly affect the
performance of the algorithm. We restrict our attention to the choice
of a factorised likelihood. Suppose that
\[
l\left(\theta\right)=\prod_{s=1}^{T}l_{s}\left(\theta\right),
\]
for some sequence $l_{s}\left(\theta\right)$. We use the notation
\[
l_{1:t}\left(\theta\right)=\prod_{s=1}^{t}l_{s}\left(\theta\right),
\]
so that $l_{1:T}\left(\theta\right)=l\left(\theta\right)$. We choose
$\pi_{t}\left(\theta\right)=p\left(\theta\right)l_{1:t}\left(\theta\right)/Z_{t}$,
and $\pi_{0}\left(\theta\right)=p\left(\theta\right)$, which encompasses
the data point tempering of \citet{Chopin2002} (where the factors
each depend on a different data point), and also the widely-used annealing
approach (where $l\left(\theta\right)=\prod_{s=1}^{T}l^{\eta_{t}-\eta_{t-1}}\left(\theta\right)$
for $\eta_{0}=0<...<\eta_{T}=1$).

The efficiency of the algorithms is also affected by the choice of
$K_{t}$ and $L_{t-1}$: here we restrict our choice to $K_{t}$ being
an MCMC move with invariant distribution $\pi_{t}$ and $L_{t-1}$
to be the reversed Markov kernel associated with $K_{t}$. When the
MCMC move mixes well, this can be an effective choice for $K_{t}$,
and its reverse is close to the optimal choice of $L_{t-1}$ when
$\pi_{t-1}$ is close to $\pi_{t}$ \citep{DelMoral2006c}. When using
an MCMC move, each iteration of the SMC algorithm will consist of
a reweighting step, a resampling step and a move step (where the MCMC
move is used). Throughout the paper we use stratified resampling \citep{douc_comparison_2005}.

In all of the SMC samplers in this section, we may use the current
set of weighted particles to inform subsequent steps of the algorithm.
In particular we describe how to use the current particles to adapt
the proposal on the active variables $q_{a}$ and the proposal on
the active variables $q_{i}$. In full generality, we may: use an
adaptive approach to resampling where the particle population is resampled
and moved when the effective sample size \citep{Kong1994} falls below
$0.5\times N_{a}$; adapt the sequence of distributions, as originally
described in \citet{DelMoral2012g} (see \citet{kerama_rare_2022}
for a description of how to implement this approach in the SMC$^{2}$
setting); and adapt the number of MCMC moves (as in \citet{South2016}).
However we do not make use of these approaches in this paper when
implementing the new approaches.

\subsection{SMC sampler with a fixed active subspace\label{subsec:SMC-sampler-with}}

\subsubsection{Construction}

In this section we describe the SMC counterpart of the AS-MH algorithm;
we embed the AS-MH move within an SMC sampler with an annealed sequence
of distributions. The structure of this algorithm has some elements
that are similar to SMC$^{2}$ \citep{Chopin2013}. We use an AS,
given by $A$ and $I$ that is fixed across the SMC iterations. Our
aim is to use an SMC sampler to simulate from the target distribution
$\pi_{a,i}$ in equation (\ref{eq:as_posterior}). From hereon we
abuse notation and denote this simply by $\pi$ to avoid additional
subscripts. The target distribution $\pi_{t}$ used at the $t$th
iteration is
\begin{eqnarray}
\pi_{t}\left(a,\left\{ i^{n}\right\} _{n=1}^{N_{i}}\right) & := & \frac{1}{Z_{t}}p_{a}\left(a\right)\prod_{j=1}^{N_{i}}q_{t,i}\left(i^{j}\mid a\right)\left(\frac{1}{N_{i}}\sum_{n=1}^{N_{i}}\frac{p_{i}\left(i^{n}\mid a\right)l_{1:t}\left(Aa+Ii^{n}\right)}{q_{t,i}\left(i^{n}\mid a\right)}\right)\label{eq:fixed_as_smc_target}
\end{eqnarray}
where $q_{t,i}$ is the proposal used for the inactive variables at
iteration $t$ and where $Z_{t}$ is a normalising constant, which
by the unbiasedness of the estimator in equation (\ref{eq:llhd_estimator})
we know to be the same as the $Z_{t}$ defined in section \ref{subsec:SMC-sampler-recap}.
We may rearrange this target as follows:

\begin{eqnarray}
\pi_{t}\left(a,\left\{ i^{n}\right\} _{n=1}^{N_{i}}\right) & = & \frac{1}{Z_{t}}p_{a}\left(a\right)\prod_{j=1}^{N_{i}}q_{t,i}\left(i^{j}\mid a\right)\left(\frac{1}{N_{i}}\sum_{n=1}^{N_{i}}\frac{p_{i}\left(i^{n}\mid a\right)l_{1:t}\left(Aa+Ii^{n}\right)}{q_{t,i}\left(i^{n}\mid a\right)}\right)\nonumber \\
 & = & \frac{1}{N_{i}}\sum_{n=1}^{N_{i}}\frac{p_{a}\left(a\right)p_{i}\left(i^{n}\mid a\right)l_{1:t}\left(Aa+Ii^{n}\right)}{Z_{t}}\left(\prod_{\substack{j=1\\
j\neq n
}
}^{N_{i}}q_{t,i}\left(i^{j}\mid a\right)\right)\nonumber \\
 & = & \frac{\pi_{t,a}\left(a\right)}{N_{i}}\sum_{n=1}^{N_{i}}\pi_{t,i}\left(i^{n}\mid a\right)\left(\prod_{\substack{j=1\\
j\neq n
}
}^{N_{i}}q_{t,i}\left(i^{j}\mid a\right)\right)\label{eq:smc_fixed_target}
\end{eqnarray}
where we used the result
\[
p_{a}\left(a\right)p_{i}\left(i\mid a\right)l_{1:t}\left(Aa+Ii^{n}\right)=Z_{t}\pi_{t,a}\left(a\right)\text{\ensuremath{\pi}}_{t,i}\left(i\mid a\right),
\]
where $\text{\ensuremath{\pi}}_{t,i}$ is the conditional distribution
defined by $\text{\ensuremath{\pi}}_{t,i}\left(i\mid a\right)=\pi_{t}\left(a,i\right)/\pi_{t,a}\left(a\right)$.
Equation (\ref{eq:smc_fixed_target}) includes in its marginals the
target at iteration $t$, $\pi_{t}\left(a,i\right)$ (here notation
is abused since we use $\pi_{t}$ to denote both the target in equation
(\ref{eq:fixed_as_smc_target}) and this marginal). The sequence of
likelihoods chosen in the previous section has the result that at
iteration $T$ of the SMC, the marginal $\pi_{T}\left(a,i\right)$
is equal to the desired posterior in equation \ref{eq:as_posterior}.

\subsubsection{Algorithm\label{subsec:Algorithm}}

In this section we use $a_{t}^{m}$ to denote the $m$th active variable
$a$-particle at the $t$th iteration of the SMC, and $i_{t}^{m,n}$
to denote the $n$th inactive variable $i$-particle for the $m$th
$a$-particle at the $t$th iteration of the SMC. Using this sequence
of targets, taking the backwards kernel to be the reverse of $K_{t},$
following \citet{DelMoral2006c} we obtain the weight update at iteration
$t$ to be

\begin{eqnarray*}
\tilde{\omega}_{t}^{m} & = & \omega_{t-1}^{m}\frac{\pi_{t}\left(a_{t-1}^{m},\left\{ i_{t-1}^{n,m}\right\} _{n=1}^{N_{i}}\right)}{\pi_{t-1}\left(a_{t-1}^{m},\left\{ i_{t-1}^{n,m}\right\} _{n=1}^{N_{i}}\right)}\\
 & = & \omega_{t-1}^{m}\frac{p_{a}\left(a_{t-1}^{m}\right)\prod_{j=1}^{N_{i}}q_{t,i}\left(i_{t-1}^{n,m}\mid a_{t-1}^{m}\right)\frac{1}{N_{i}}\sum_{n=1}^{N_{i}}\tilde{w}_{t}^{n,m}\left(a_{t-1}^{m},i_{t-1}^{n,m}\right)}{p_{a}\left(a_{t-1}^{m}\right)\prod_{j=1}^{N_{i}}q_{t-1,i}\left(i_{t-1}^{n,m}\mid a_{t-1}^{m}\right)\frac{1}{N_{i}}\sum_{n=1}^{N_{i}}\tilde{w}_{t-1}^{n,m}\left(a_{t-1}^{m},i_{t-1}^{n,m}\right)}\\
 & = & \omega_{t-1}^{m}\frac{\prod_{j=1}^{N_{i}}q_{t,i}\left(i_{t-1}^{n,m}\mid a_{t-1}^{m}\right)\sum_{n=1}^{N_{i}}\tilde{w}_{t}^{n,m}\left(a_{t-1}^{m},i_{t-1}^{n,m}\right)}{\prod_{j=1}^{N_{i}}q_{t-1,i}\left(i_{t-1}^{n,m}\mid a_{t-1}^{m}\right)\sum_{n=1}^{N_{i}}\tilde{w}_{t-1}^{n,m}\left(a_{t-1}^{m},i_{t-1}^{n,m}\right)},
\end{eqnarray*}
where
\begin{equation}
\tilde{w}_{t}^{n,m}\left(a_{t-1}^{m},i_{t-1}^{n,m}\right)=\frac{p_{i}\left(i_{t-1}^{n,m}\mid a_{t-1}^{m}\right)l_{1:t}\left(Aa_{t-1}^{m}+Ii_{t-1}^{n}\right)}{q_{t,i}\left(i_{t-1}^{n,m}\mid a_{t-1}^{m}\right)},\label{eq:unnorm_w}
\end{equation}
with the $t$ subscript for the unnormalised weight $\tilde{w}_{t}^{n,m}$
denoting that it uses the likelihood $l_{1:t}$ and proposal $q_{t,i}$
(so that $\tilde{w}_{t-1}^{n,m}$ follows equation (\ref{eq:unnorm_w}),
but uses $l_{1:t-1}$ and $q_{t-1,i}$). At iteration $t$, for each
particle we use an MCMC move with invariant distribution $\pi_{t}\left(a,\left\{ i^{n}\right\} _{n=1}^{N_{i}}\right)$;
for this we may use an AS-MH move with likelihood $l_{1:t}$. The
active subspace SMC (AS-SMC) algorithm is given in algorithm \ref{alg:as_smc}.
Discussion of how to tune the proposals $q_{t,a}$ and $q_{t,i}$
is found in sections \ref{subsec:MH-on} and \ref{subsec:Choice-of}
respectively.

\begin{algorithm}
\caption{AS-SMC}\label{alg:as_smc}
\small

Simulate $N_{a}$ points, $\left\{ \theta_{0}^{m} \right\}_{m=1}^{N_{a}} \sim p$ and set each weight $\omega_0^m = 1/N_{a}$;\

Find AS using points $\left\{ \theta_{0}^{m} \right\}_{m=1}^{N_{a}}$ in equation (\ref{eq:est_as}), yielding $A$ and $I$;\

\For{$m=1:N_{a}$}
{
	Set $a^m_{0}=A^{T}\left(\theta^m_{0}\right)$, $i^{1,m}_{0}=I^{T}\left(\theta^m_{0}\right)$ and $u_0^m = 1$;\

	\For{$n=2:N_{i}$}
	{         
		$i_{0}^{n,m} \sim q_{0,i}\left( \cdot \mid a_{0}^m \right) := p_i\left( \cdot \mid a_{0}^m \right)$;\
	}

	Set $w_0^{n,m} = 1/N_{i}$ for $n=1:N_i$;\
}

\For {$t=1:T$}
{
	\For(\tcp*[h]{reweight}){$m=1:N_{a}$}
	{
			\For{$n=1:N_{i}$}
			{         
				$\tilde{w}_{t}^{n,m}\left(a_{t-1}^{m},i_{t-1}^{n,m}\right)=\frac{p_{i}\left(i_{t-1}^{n,m}\mid a_{t-1}^{m}\right)l_{1:t}\left(B_{a}a_{t-1}^{m}+B_{i}i_{t-1}^{n,m}\right)}{q_{t,i}\left(i_{t-1}^{n,m}\mid a_{t-1}^{m}\right)}$;\
			}
			
			\[
			\tilde{\omega}_{t}^{m} = \omega_{t-1}^{m}\frac{\prod_{j=1}^{N_{i}}q_{t,i}\left(i_{t-1}^{n,m}\mid a_{t-1}^{m}\right)\sum_{n=1}^{N_{i}}\tilde{w}_{t}^{n,m}\left(a_{t-1}^{m},i_{t-1}^{n,m}\right)}{\prod_{j=1}^{N_{i}}q_{t-1,i}\left(i_{t-1}^{n,m}\mid a_{t-1}^{m}\right)\sum_{n=1}^{N_{i}}\tilde{w}_{t-1}^{n,m}\left(a_{t-1}^{m},i_{t-1}^{n,m}\right)};
			\]
	}

	$\left\{ \omega^m_{t} \right\}_{m=1}^{N_{a}} \leftarrow \mbox{ normalise}\left( \left\{ \tilde{\omega}^m_{t} \right\}_{m=1}^{N_{a}} \right)$;

	\For{$m=1:N_{a}$}
	{
		$\left\{ w^{n,m}_{t} \right\}_{n=1}^{N_{i}} \leftarrow \mbox{ normalise}\left( \left\{ \tilde{w}^{n,m}_{t} \right\}_{n=1}^{N_{i}} \right)$;
	}

	\For{$m=1:N_{a}$}
	{
		$u_t^{m} \sim\mathcal{M}\left( \left( w_t^{1,m}, ..., w_t^{N_i,m} \right) \right)$;
	}

	\If(\tcp*[h]{resample and move}) {some degeneracy condition is met}
	{
		\For{$m=1:N_{a}$}
		{
			Simulate $\left(a_{t}^{m}, i_{t}^{1:N_{i},m} \right)$ from the mixture distribution
			\[
			\sum_{j=1}^{N_{a}} \omega_t^{j} K_{t,a}\left\{ \cdot \mid \left(a_{t-1}^{j},i_{t-1}^{1:N_{i},j}\right) \right\},
			\]
			where $K_{t,a}$ is an AS-MH move, i.e.:

			$j^* \sim \mathcal{M}\left( \left\{ \omega_t^{j} \right\}_{j=1}^{N_{a}} \right)$; $ a_{t}^{*m} \sim q_a\left(\cdot\mid a_{t}^{j^*} \right)$;\

			\For {$n=1:N_i$}
			{
		
				$i_{t}^{*n,m} \sim q_{t,i} \left(\cdot \mid a_{t}^{*m} \right);$\
		
				\[
				\tilde{w}_t^{n,m}\left(a_{t}^{*m},i_{t,i}^{*n,m}\right) = \frac{p_{i}\left(i^{*n,m}_{t} \mid a^{*m}_{t}\right) l_{1:t}\left( A a_{t}^{*m} + Ii_{t}^{*n,m}\right)}{q_{t,i}\left(i_{t}^{*n,m}\mid a_{t}^{*m} \right)};
				\]\
			}
		
			$u_t^{*m} \sim\mathcal{M}\left( \left( w_t^{*1,m}, ..., w_t^{*N_i,m} \right) \right),$ where for $n=1:N_i$
			\[
			w_t^{*n,m} = \frac{\tilde{w}_t^{n,m}\left(a_{t}^{*m},i_{t}^{*n,m}\right)}{\sum_{p=1}^{N_{i}} \tilde{w}_t^{p,m}\left(a_{t}^{*m},i_{t}^{*p,m}\right)};\
			\]
		
		   Set $\left(a_{t}^m, \left\{ i_{t}^{n,m}, \tilde{w}_t^{n,m} \right\}_{n=1}^{N_i}, u_t^m \right) = \left(a_{t}^{*m},\left\{ i_{t}^{*n,m}, \tilde{w}_t^{*n,m} \right\}_{n=1}^{N_i}, u_t^{*m} \right)$ with probability
			\[
			\alpha_{t,a}^m = 1\wedge\frac{p_{a}\left(a^{*m}_{a}\right) \sum_{n=1}^{N_{i}}\tilde{w}_{t}^{n,m}\left(a_{t}^{*m},i_{t}^{*n,m}\right)}{p_{a}\left(a^{j^*}_{t}\right) \sum_{n=1}^{N_{i}}\tilde{w}_{t}^{n,j^*}\left(a_{t-1}^{j^*},i_{t-1}^{n,j^*}\right)}\frac{q_{t,a}\left(a_{t}^{j^*} \mid a_{t}^{*m}\right)}{q_{t,a}\left(a_{t}^{*m}\mid a_{t}^{j^*} \right)};
			\]\
			
			Else let $\left(a_{t}^m, \left\{ i_{t}^{n,m}, \tilde{w}_t^{n,m} \right\}_{n=1}^{N_i}, u_t^m \right) = \left(a_{t}^{j^*},\left\{ i_{t}^{n,j^*}, \tilde{w}_t^{n,j^*} \right\}_{n=1}^{N_i}, u_t^{j^*}\right)$;\
		}

		$\omega^m_{t} = 1/N_{a}$ for $m=1:N_{a}$;
	}

}
\end{algorithm}

\citet{Chopin2013} shows, for estimating the expectation of $g$
with respect to $\pi_{t}$, we may use either of the following
\begin{enumerate}
\item \textbf{Using one $i$-point for each $a$-point:}
\begin{equation}
\sum_{m=1}^{N_{a}}\omega_{t}^{m}g\left(Aa_{t}^{m}+Ii_{t}^{u_{t}^{m},m}\right).\label{eq:est1_smc}
\end{equation}
\item \textbf{Using all of the accepted $i$-points for each $a$-point:}
\begin{equation}
\sum_{m=1}^{N_{a}}\omega_{t}^{m}\sum_{n=1}^{N_{i}}w_{t}^{n,m}g\left(Aa_{t}^{m}+Ii_{t}^{n,m}\right).\label{eq:est2_smc}
\end{equation}
\end{enumerate}
To see that the first of these holds, we consider an extended version
of the target distribution in equation \ref{eq:smc_fixed_target},
as we will see in section \ref{subsec:SMC-sampler-with-1}.

\subsection{SMC sampler with an adaptive active subspace\label{subsec:SMC-sampler-with-1}}

We now examine the case of using an AS, where the matrices now denoted
as $B_{a_{t}}$ and $B_{i_{t}}$ change with $t$, across SMC iterations.
At the $t$th iteration of the algorithm, $A_{t}$ and $I_{t}$ are
determined by using the current particle population\@. The change
of space is implemented at the beginning of each SMC iteration, as
in the \emph{transformation SMC} of \citet{everitt_sequential_2020}.

\subsubsection{Adapting the active subspace}

At the beginning of each SMC iteration, we find an estimate of the
AS that we will use at that iteration. At the beginning of the $t$th
iteration we estimate the matrix in equation (\ref{eq:as}) for the
score function $\nabla\log l_{1:t}$ using $\phi=\pi_{t-1}$, with
equation (\ref{eq:est2_smc})
\begin{equation}
\sum_{m=1}^{N_{a}}\omega_{t-1}^{m}\sum_{n=1}^{N_{i}}w_{t-1}^{n,m}\nabla\log l_{1:t}\left(Aa_{t-1}^{m}+Ii_{t-1}^{n,m}\right)\nabla\log l_{1:t}\left(Aa_{t-1}^{m}+Ii_{t-1}^{n,m}\right)^{T}.\label{eq:smc_as}
\end{equation}
The eigendecomposition, together with our chosen approach to choose
the active and inactive variables, yields the AS that we represent
using $A_{t}$ and $I_{t}$.

The next step is that we need to find a representation of the current
particle set in the new AS. To do this, for each particle we select
a single inactive variable, then transform the active variable and
the selected inactive variable from iteration $t$ to the original
space of $\theta$, then project the resultant $\theta$ into the
new AS. We then require an additional $N_{i}-1$ inactive variables
for estimating the marginal likelihood. This move needs to be carefully
constructed: we use a conditional IS update \citep{Andrieu2010c,Chopin2013}.

We allow for the possibility for the dimension $d_{t,i}$ of the inactive
subspace to reduce to zero, but we require that the dimension $d_{t,a}$
of the AS is always greater than zero. When $d_{t,i}=0$, we no longer
require an estimated likelihood and the target is simplified significantly.

\subsubsection{Target distributions\label{subsec:Construction}}

This algorithm uses the target
\begin{eqnarray}
\pi_{t}\left(a_{t},\left\{ i_{t}^{n}\right\} _{n=1}^{N_{i}}\right) & := & \frac{1}{Z_{t}}p_{t,a}\left(a_{t}\right)\prod_{j=1}^{N_{i}}q_{t,i}\left(i_{t}^{j}\mid a_{t}\right)\frac{1}{N_{i}}\sum_{n=1}^{N_{i}}\frac{p_{t,i}\left(i_{t}^{n}\mid a_{t}\right)l_{1:t}\left(A_{t}a_{t}+i_{t}i_{t}^{n}\right)}{q_{t,i}\left(i_{t}^{n}\mid a_{t}\right)}\label{eq:adaptive_as_smc_target}\\
 & = & \frac{\pi_{t,a}\left(a_{t}\right)}{N_{i}}\sum_{n=1}^{N_{i}}\pi_{t,i}\left(i_{t}^{n}\mid a_{t}\right)\left(\prod_{\substack{j=1\\
j\neq n
}
}^{N_{i}}q_{t,i}\left(i_{t}^{j}\mid a_{t}\right)\right)\label{eq:adaptive_as_smc_target_alt}
\end{eqnarray}
at iteration $t$, where $p_{t,a}$ and $\pi_{t,a}$ denote the marginal
prior and posterior of $a$ under the AS determined at iteration $t$
and $p_{t,i}$ and $\pi_{t,i}$ denote the corresponding conditional
prior and posterior on $i\mid a$. Since the AS has changed between
iterations, to easily describe the algorithm we additionally need
to define a target that uses likelihood $l_{1:t-1}$, but which uses
the AS at iteration $t$. This is given by
\begin{eqnarray}
\lambda_{t-1}\left(a_{t},\left\{ i_{t}^{n}\right\} _{n=1}^{N_{i}}\right) & := & \frac{1}{Z_{t-1}}p_{t,a}\left(a_{t}\right)\prod_{j=1}^{N_{i}}\kappa_{t-1,i}\left(i_{t}^{j}\mid a_{t}\right)\frac{1}{N_{i}}\sum_{n=1}^{N_{i}}\frac{p_{t,i}\left(i_{t}^{n}\mid a_{t}\right)l_{1:t-1}\left(A_{t}a_{t}+I_{t}i_{t}^{n}\right)}{\kappa_{t-1,i}\left(i_{t}^{n}\mid a_{t}\right)}\label{eq:adaptive_as_smc_target-1}\\
 & = & \frac{\lambda_{t-1,a}\left(a_{t}\right)}{N_{i}}\sum_{n=1}^{N_{i}}\lambda_{t-1,i}\left(i_{t}^{n}\mid a_{t}\right)\left(\prod_{\substack{j=1\\
j\neq n
}
}^{N_{i}}\kappa_{t-1,i}\left(i_{t}^{j}\mid a_{t}\right)\right),\label{eq:adaptive_as_smc_target_alt-1}
\end{eqnarray}
where $\lambda_{t-1,a}$ and $\lambda_{t-1,i}$ denote the posterior
using $l_{1:t-1}$, projected onto the AS from iteration $t$. For
simplicity we take $\kappa_{t-1,i}$ to be equal to $q_{t,i}$ (in
practice both will be taken to be $p_{t,i}$), although in full generality
these two proposals could be chosen independently. The SMC then uses
two steps at iteration $t$: the first a conditional IS step to move
from $\pi_{t-1}$ to $\lambda_{t-1}$ which we call `reprojection';
the second to move from $\pi_{t-1\rightarrow t}$ to $\pi_{t}$ which
is a standard weight update. The following two sections detail these
steps. Following these steps, we resample, then for each particle
use an MCMC move with invariant distribution $\pi_{t}\left(a_{t},\left\{ i_{t}^{n}\right\} _{n=1}^{N_{i}}\right)$,
using an AS-MH move with likelihood $l_{1:t}$.

\subsubsection{Reprojection}

At the beginning of each iteration, an eigendecomposition of equation
(\ref{eq:smc_as}) is used to determine $A_{t}$ and $I_{t}$. We
then perform the following conditional IS step on each particle to
obtain a particle in the new active and inactive subspaces, moving
particle $\left(a_{t-1}^{m},\left\{ i_{t-1}^{n,m}\right\} _{n=1}^{N_{i}}\right)$
to $\left(a_{t-1\rightarrow t}^{m},\left\{ i_{t-1\rightarrow t}^{n,m}\right\} _{n=1}^{N_{i}}\right)$.
We cannot easily transform the first collection of weighted points
to the second, since we have multiple inactive points per each active
point each with its own (internal) weight. Instead we need to select
one of the inactive points to pair with the active, transform back
to the original $\theta$-space, then transform to the new parameterisation.
This involves discarding all but one of the inactive points, then
regenerating additional inactive points after the reparameterisation.
To construct this move we use the observation \citet{Chopin2013}
that the target in equation (\ref{eq:adaptive_as_smc_target_alt})
can be seen as a marginalisation of an extended distribution $\pi_{t}^{*}$
over a uniformly distributed particle index variable $u_{t}$
\begin{equation}
\pi_{t}^{*}\left(u_{t},a_{t},\left\{ i_{t}^{n}\right\} _{n=1}^{N_{i}}\right)=\frac{\pi_{t,a}\left(a_{t}\right)}{N_{i}}\pi_{t,i}\left(i_{t}^{u_{t}}\mid a_{t}\right)\left(\prod_{\substack{j=1\\
j\neq u_{t}
}
}^{N_{i}}q_{t,i}\left(i_{t}^{j}\mid a_{t}\right)\right).\label{eq:adaptive_smc_extended_target}
\end{equation}
\citet{Chopin2013} notes that it is simple to extend a set of weighted
particles from $\pi_{t-1}$ so that they are from $\pi_{t-1}^{*}$:
for each particle we simulate from the conditional distribution of
$u_{t-1}$, which is given by $u_{t-1}\mid a_{t-1},\left\{ i_{t-1}^{n}\right\} _{n=1}^{N_{i}}\sim\mathcal{M}\left(\left(w_{t-1}^{1,m},...,w_{t-1}^{N_{i},m}\right)\right)$,
where $w_{t-1}^{n,m}$ is the normalised version of $\tilde{w}_{t-1}^{n,m}\left(a_{t-1}^{m},i_{t-1}^{n,m}\right)$.
At the beginning of iteration $t$, our method performs this simulation
of $u_{t-1}$ for each particle, then makes use of the following transformation
of the extended state
\[
a_{t-1\rightarrow t}:=G_{t-1\rightarrow t,a}\left(u_{t-1},a_{t-1},\left\{ i_{t-1}^{n}\right\} _{n=1}^{N_{i}}\right):=A_{t}^{T}\left(A_{t-1}a_{t-1}+I_{t-1}i_{t-1}^{u_{t-1}}\right)
\]
\[
i_{t-1\rightarrow t}^{n}:=G_{t-1\rightarrow t,i}\left(u_{t-1},a_{t-1},\left\{ i_{t-1}^{n}\right\} _{n=1}^{N_{i}}\right):=I_{t}^{T}\left(A_{t-1}a_{t-1}+I_{t-1}i_{t-1}^{n}\right)
\]
The conditional IS move makes use of this transformation, along with
a proposal from \citet{Chopin2013}. Our desired target distribution
for the new point is $\lambda_{t-1}^{*}\left(u_{t-1},a_{t-1\rightarrow t},\left\{ i_{t-1\rightarrow t}^{n}\right\} _{n=1}^{N_{i}}\right)$,
the extension of the target in equation (\ref{eq:adaptive_as_smc_target-1}),
just as equation (\ref{eq:adaptive_smc_extended_target}) is an extension
of (\ref{eq:adaptive_as_smc_target}). The 

Our proposal uses the current particle with values $\left(u_{t-1},a_{t-1},\left\{ i_{t-1}^{n}\right\} _{n=1}^{N_{i}}\right)$
passed through the transformation above to give the point $\left(u_{t-1},a_{t-1\rightarrow t},i_{t-1\rightarrow t}^{u_{t-1}}\right)$,
plus the variables $\left\{ i_{t-1\rightarrow t}^{n}\right\} _{n=1,n\neq u_{t-1}}^{N_{i}}$
which will be discarded. We then let $a_{t}=a_{t-1\rightarrow t}$
and $i_{t}^{u_{t-1}}=i_{t-1\rightarrow t}^{u_{t-1}}$. In place of
the discarded variables we additionally require the variables $\left\{ i_{t}^{n}\right\} _{n=1,n\neq u_{t-1}}^{N_{i}}$,
proposing them from the conditional distribution on the inactive variables
resulting from $\lambda_{t-1}^{*}$. Following the structure of equation
(\ref{eq:adaptive_smc_extended_target}) and using equation (\ref{eq:adaptive_as_smc_target_alt-1}),
this conditional distribution is given by
\begin{eqnarray*}
\frac{\lambda_{t-1}^{*}\left(u,a_{t},\left\{ i_{t}^{n}\right\} _{n=1}^{N_{i}}\right)}{\frac{\lambda_{t-1,a}\left(a_{t}\right)}{N_{i}}\lambda_{t-1,i}\left(i_{t}^{u}\mid a_{t}\right)} & = & \prod_{\substack{j=1\\
j\neq u
}
}^{N_{i}}\kappa_{t-1,i}\left(i_{t}^{j}\mid a_{t}\right),
\end{eqnarray*}
For the IS to be valid, we need to artificially extend the target
distribution using a backwards kernel $L$ (as in \citet{DelMoral2006c})
to form a joint distribution over all of the variables involved in
the proposal, such that the desired target $\pi_{t}^{*}$ is a marginal
distribution. The IS target is then
\[
\lambda_{t-1}^{*}\left(u,a_{t},\left\{ i_{t}^{n}\right\} _{n=1}^{N_{i}}\right)L\left(\left\{ i_{t-1\rightarrow t}^{n}\right\} _{n=1,n\neq u}^{N_{i}}\mid u,a_{t},\left\{ i_{t}^{n}\right\} _{n=1}^{N_{i}}\right)
\]
and the proposal is
\[
\pi_{t-1\rightarrow t}^{*}\left(u,a_{t},i_{t}^{n},\left\{ i_{t-1\rightarrow t}^{n}\right\} _{n=1,n\neq u}^{N_{i}}\right)\text{\ensuremath{\prod_{\substack{j=1\\
j\neq u
}
}^{N_{i}}\kappa_{t-1,i}\left(i_{t}^{j}\mid a_{t}\right)}},
\]
where $\pi_{t-1\rightarrow t}^{*}$ pushforward distribution of $\pi_{t-1}^{*}$
under $G_{t-1\rightarrow t}$. We choose the backwards kernel to be

\[
L\left(\left\{ i_{t-1\rightarrow t}^{n}\right\} _{n=1,n\neq u}^{N_{i}}\mid u,a_{t},\left\{ i_{t}^{n}\right\} _{n=1}^{N_{i}}\right)=\frac{N_{i}\pi_{t-1\rightarrow t}^{*}\left(u,a_{t},i_{t}^{n},\left\{ i_{t-1\rightarrow t}^{n}\right\} _{n=1,n\neq u}^{N_{i}}\right)}{\lambda_{t-1,a}\left(a_{t}\right)\lambda_{t-1,i}\left(i_{t}^{u}\mid a_{t}\right)},
\]
which gives an importance weight of 1 (as in \citet{Chopin2013}).
To proceed with the next iteration of the SMC, we then discard the
value $u_{t-1}$, so that our particle is from the marginal distribution
$\pi_{t}$, rather than the extended target $\pi_{t}^{*}$.

In summary:
\begin{itemize}
\item this additional step is run after determining the AS for the next
iteration, for each of the $N_{a}$ particles;
\item we sample one of the inactive particles, using the weights of the
particles in the inactive space;
\item we project the active variable and sampled inactive variable into
the new active and inactive subspaces;
\item we sample $N_{i}-1$ additional inactive variables from the proposal
$\kappa_{t-1,i}$.
\end{itemize}

\subsubsection{Weight update}

The weight update at iteration $t$ is similar to that in section
\ref{subsec:SMC-sampler-with}
\begin{eqnarray*}
\tilde{\omega}_{t}^{m} & = & \omega_{t-1}^{m}\frac{\pi_{t}\left(a_{t-1\rightarrow t}^{m},\left\{ i_{t-1\rightarrow t}^{n,m}\right\} _{n=1}^{N_{i}}\right)}{\pi_{t-1\rightarrow t}\left(a_{t-1\rightarrow t}^{m},\left\{ i_{t-1\rightarrow t}^{n,m}\right\} _{n=1}^{N_{i}}\right)}\\
 & = & \omega_{t-1}^{m}\frac{\prod_{j=1}^{N_{i}}q_{t,i}\left(\theta_{i_{t}}^{j}\mid\theta_{a_{t}}\right)\sum_{n=1}^{N_{i}}\tilde{w}_{t}^{n,m}\left(a_{t-1\rightarrow t}^{m},i_{t-1\rightarrow t}^{n,m}\right)}{\prod_{j=1}^{N_{i}}q_{t-1\rightarrow t,i}\left(\theta_{i_{t}}^{j}\mid\theta_{a_{t}}\right)\sum_{n=1}^{N_{i}}\tilde{w}_{t-1}^{n,m}\left(a_{t-1\rightarrow t}^{m},i_{t-1\rightarrow t}^{n,m}\right)},
\end{eqnarray*}
where
\[
\tilde{w}_{t}^{n,m}\left(a_{t-1\rightarrow t}^{m},i_{t-1\rightarrow t}^{n,m}\right)=\frac{p_{i}\left(i_{t-1\rightarrow t}^{n,m}\mid a_{t-1\rightarrow t}^{m}\right)l_{1:t}\left(A_{t}a_{t-1\rightarrow t}^{m}+I_{t}i_{t-1\rightarrow t}^{n}\right)}{q_{t,i}\left(i_{t-1\rightarrow t}^{n,m}\mid a_{t-1\rightarrow t}^{m}\right)},
\]
similarly to section \ref{subsec:SMC-sampler-with} and
\[
\tilde{w}_{t-1}^{n,m}\left(a_{t-1\rightarrow t}^{m},i_{t-1\rightarrow t}^{n,m}\right)=\frac{p_{i_{t}}\left(i_{t-1\rightarrow t}^{n,m}\mid a_{t-1\rightarrow t}^{m}\right)l_{1:t-1}\left(A_{t}a_{t-1\rightarrow t}^{m}+I_{t}i_{t-1\rightarrow t}^{n}\right)}{q_{t-1\rightarrow t,i}\left(i_{t-1\rightarrow t}^{n,m}\mid a_{t-1\rightarrow t}^{m}\right)}.
\]

The full algorithm, adaptive AS-SMC is given in algorithm \ref{alg:adaptive_as_smc}.

\subsubsection{MH on $a$\label{subsec:MH-on}}

The use of an AS means that we expect our moves on $a$ to be more
`difficult' than those on $i$. To ensure the MCMC is efficient,
we may require moves on $a$ that use the gradient of the posterior
$\pi_{t,a}\left(a\right)$, such as the Metropolis-adjusted Langevin
algorithm (MALA). In the pseudo-marginal case this gradient is not
available analytically, but it may be estimated as in, e.g. \citet{dahlin_particle_2015}.
For simplicity we do not consider this approach in the paper; instead
we use a Gaussian random walk proposal. At iteration $t$ we take
$q_{t,a}$ to be a multivariate Gaussian with mean the current point
and covariance $\Sigma_{t,q_{a}}$. Based on \citet{sherlock_random_2010},
at iteration $t$ we take $\Sigma_{t,q_{a}}$ to be the scaled empirical
covariance of the reprojected particles from iteration $t-1$
\[
\Sigma_{q_{a}}=\hat{\Sigma}_{t-1\rightarrow t,a}=\frac{2.38^{2}}{d_{a}}\sum_{m=1}^{N_{a}}\omega_{t}^{m}\left(a_{t-1\rightarrow t}^{m}-\hat{\mu}_{t-1\rightarrow t,a}\right)\left(a_{t-1\rightarrow t}^{m}-\hat{\mu}_{t-1\rightarrow t,a}\right)^{T}
\]
where
\[
\hat{\mu}_{t-1\rightarrow t,a}=\sum_{m=1}^{N_{a}}\omega_{t}^{m}a_{t-1\rightarrow t}^{m}.
\]
This adaptation takes place directly following line 20 of algorithm
\ref{alg:as_smc}, or after line 28 of algorithm \ref{alg:adaptive_as_smc}.

\subsubsection{Choice of $q_{t,i}$\label{subsec:Choice-of}}

For the moves on $i$, we use a proposal $q_{t,i}$. In the case of
an ideal AS $q_{t,i}\left(i\mid a\right)=p_{i}\left(i\mid a\right)$
is guaranteed to be a good choice, but outside of this situation we
wish to adapt $q_{t,i}$ using the particle population. For this paper
we propose two possible choices for the proposal $q_{t,i}$ at iteration
$t$. One is to use a multivariate Gaussian $\mathcal{MVN}\left(\mu_{q_{i}},\Sigma_{q_{i}}\right)$;
the other is to use a multivariate $t$-distribution with 5 degrees
of freedom $t_{5}\left(\mu_{q_{i}},\Sigma_{q_{i}}\right)$ (as in
\citet{sherlock_random_2010}). The former would be close to coinciding
with the choice $q_{t,i}\left(\theta_{i}\mid\theta_{a}\right)=p_{i}\left(\theta_{i}\mid\theta_{a}\right)$
in the common case where the prior is chosen to be Gaussian; the latter
is theoretically more likely to result in a more effective sampler
in general \citep{mengersen_rates_1996}.

We propose to adapt the parameters $\mu_{t,q_{i}},\Sigma_{t,q_{i}}$
following line 10 of algorithm \ref{alg:as_smc}, or after line 17
of algorithm \ref{alg:adaptive_as_smc}. The optimal choice of $q_{t,i}\left(i\mid a\right)$
would be proportional to $p_{i}\left(i\mid a\right)l_{1:t}\left(A_{t}a+I_{t}i\right)$.
From the previous iteration, for each $a^{m}$, we have a population
of $N_{i}-1$ weighted points distributed according to $q_{t-1,i}\left(\cdot\mid a^{m}\right)$.
We use these populations to approximate the mean and covariance of
the optimal choice using importance sampling. We first calculate the
weight
\[
\tilde{\kappa}_{t}^{n,m}=w_{t-1}^{n,m}l_{t}\left(A_{t}a_{t-1\rightarrow t}^{m}+I_{t}i_{t-1\rightarrow t}^{n,m}\right)
\]
for $m=1:N_{a}$ and $n=1:N_{i},n\neq u_{t-1\rightarrow t}^{m}$,
then for $m=1:N_{a}$ normalise $\left(\tilde{\kappa}_{t}^{1,m},...,\tilde{\kappa}_{t}^{u_{t-1}^{m}-1,m},\tilde{\kappa}_{t}^{u_{t-1}^{m}+1,m},...,\tilde{\kappa}_{t}^{N_{i},m}\right)$
to yield normalised weights $\left(\kappa_{t}^{1,m},...,\kappa_{t}^{u_{t-1\rightarrow t}^{m}-1,m},\kappa_{t}^{u_{t-1\rightarrow t}^{m}+1,m},...,\kappa_{t}^{N_{i},m}\right)$.
For $m=1:N_{a}$ we then calculate
\[
\hat{\mu}_{t-1\rightarrow t,i}^{m}=\sum_{\substack{n=1\\
n\neq u_{t-1\rightarrow t}^{m}
}
}^{N_{i}}\kappa_{t}^{n,m}i_{t-1\rightarrow t}^{n,m}
\]

\[
\hat{\Sigma}_{t-1\rightarrow t,i}^{m}=\sum_{\substack{n=1\\
n\neq u_{t-1\rightarrow t}^{m}
}
}^{N_{i}}\kappa_{t}^{n,m}\left(i_{t-1\rightarrow t}^{n,m}-\hat{\mu}_{t-1\rightarrow t}^{m}\right)\left(i_{t-1\rightarrow t}^{n,m}-\hat{\mu}_{t-1\rightarrow t,i}^{m}\right)^{T}
\]
and use $\hat{\mu}_{t-1\rightarrow t,i}^{m}$ and $\hat{\Sigma}_{t-1\rightarrow t,i}^{m}$
as the parameters of the proposal $q_{t,i}\left(\cdot\mid\theta_{a}^{m}\right)$.
For algorithm \ref{alg:adaptive_as_smc}, we take $q_{t-1\rightarrow t,i}=p_{t,i}$.

\subsection{Active subspace SMC$^{2}$\label{subsec:SMC-variant}}

A weakness of the AS-SMC, is the same as AS-MH: that the variance
of the IS estimator of the likelihood in equation \ref{eq:llhd_estimator}
will grow exponentially as the dimension of the inactive variables
increases, even if the likelihood is only a little informative about
these variables. A way of circumventing this problem is to change
the marginal likelihood estimator from the one in equation \ref{eq:llhd_estimator}
to an SMC estimator, which is likely to have a lower variance in high
dimensions. 

We propose to estimate \ref{eq:llhd} using an SMC sampler with the
sequence of distributions $\pi_{0,i}\left(i\mid a\right)=p_{i}\left(i\mid a\right)$
and $\pi_{t,i}\left(i\mid a\right)\propto p_{i}\left(i\mid a\right)l_{1:t}\left(Aa+Ii\right)$
for $t=1:T$, assuming once again that the AS is fixed across the
SMC iterations. For the move step in this SMC, at iteration $t$ we
propose to use an MH move with independent proposal $q_{t,i}$. This
`internal' SMC sampler may be embedded within an `external' SMC
sampler on the space of $a$ using the approach of \citet{Chopin2013}.
The algorithm follows exactly the SMC$^{2}$ framework of \citet{Chopin2013}.
Full details are found in algorithms \ref{alg:inactive_smc} and \ref{alg:as_smc2}
in the appendix, where algorithm \ref{alg:inactive_smc} is called
by algorithm \ref{alg:as_smc2}. The primary difference between AS-SMC
and AS-SMC$^{2}$ are that the inactive variables are not drawn independently
at each SMC iteration. Instead the inactive variables from the previous
step are stored and reused: SMC$^{2}$ guides a set of particles to
find `good' values for the inactive variables. The justification
for this algorithm can be found in appendix \ref{sec:Justification-for-AS-SMC}.

To choose $q_{s,i}$ in algorithm \ref{alg:inactive_smc}, we use
a similar scheme to that in section \ref{subsec:Choice-of}, where
the adaptation is performed at line 10, and $\hat{\mu}_{s-1,i}$ and
$\hat{\Sigma}_{s-1,i}$ are used as the parameters of the proposal
$q_{s,i}\left(\cdot\right)$, with
\[
\hat{\mu}_{s-1,i}=\sum_{m=1}^{N_{a}}\omega_{t}^{m}\sum_{n=1}^{N_{i}}w_{s}^{n}i_{s-1}^{n}
\]

\[
\hat{\Sigma}_{s-1,i}=\sum_{m=1}^{N_{a}}\omega_{t}^{m}\sum_{n=1}^{N_{i}}w_{s}^{n}\left(i_{s-1}^{n}-\hat{\mu}_{s-1,i}\right)\left(i_{s-1}^{n}-\hat{\mu}_{s-1,i}\right)^{T}.
\]

\subsubsection{SMC$^{2}$ variant of adaptive AS-SMC}

We might also consider the SMC$^{2}$ extension in adaptive AS-SMC,
although it is significantly more complex and with less potential
gain. The reason for this is that, since the AS is changed throughout
the algorithm, the reprojection step introduced in section \ref{subsec:Construction}
is required to simulate the inactive particles at each new SMC iteration,
and this involves resampling $N_{i}-1$ particles from the proposal
distribution $q_{t,i}$. The main advantage of AS-SMC$^{2}$ is that
the set of $N_{i}$ inactive particles are maintained and refined
throughout the algorithm; an advantage which is mostly lost through
use of the reprojection step, where only one of the existing inactive
particles would be propagated.

As a substitute for this SMC$^{2}$ approach, we propose the following
alternative: that algorithm \ref{alg:adaptive_as_smc} is followed
for the initial SMC iterations then, the after some criterion is met,
AS-SMC$^{2}$ is followed for the remainder of the target distributions.
This switching criterion should be based on the estimates of the AS
stabilising, which we might expect as $t$ increases and the likelihood
becomes increasingly influential. After switching, the particles representing
the inactive variables may be replaced with particles drawn from the
SMC sampler on $\theta_{i}$-space in AS-SMC$^{2}$, with the $\theta_{a}$
particles reweighted according to the `exchange importance sampling
step' from \citet{Chopin2013}. We do not pursue this approach further
in this paper.

\section{Empirical results\label{sec:Empirical-results}}

\subsection{Comparing SMC, AS-SMC and AS-SMC$^{2}$}

\subsubsection{Models}

As an example of where active subspaces are applicable, we use a toy
Gaussian model. We perform inference for parameter $\theta=(\theta_{1},...,\theta_{d})^{T}\in\mathbb{R}^{d}$,
given observations of variable $y\in\mathbb{R}$. We use prior $\theta_{i}\sim\mathcal{N}\left(0,\sigma=\sqrt{5000}\right)$
and model $y\sim\mathcal{N}\left(\sum_{i=1}^{d}\theta_{i},1\right)$.
We fit this model to 100 observations of $y$ drawn from $\mathcal{N}(0,1)$.
This model is useful for studying the case of an `ideal' AS. The
$\theta_{i}$ can only be identified as lying on a hyperplane of dimension
$k-1$ with equation $\sum_{i=1}^{k}\theta_{i}=0$. Figure \ref{fig:A-2d-slice}
gives an illustration of this model for $d=3$. We call this model
the `plane' model.

\begin{figure}
\subfloat[\foreignlanguage{british}{A 2d slice of the Gaussian prior on the horizontal plane $\theta_{1}=0$
together with the level surface of the likelihood in the particular
case $\theta_{1}+\theta_{2}+\theta_{3}+b\theta_{2}^{2}+b\theta_{3}^{2}=0$
for $b=0.001$ (green curved plane).\label{fig:A-2d-slice-1}}]{\includegraphics[scale=0.4]{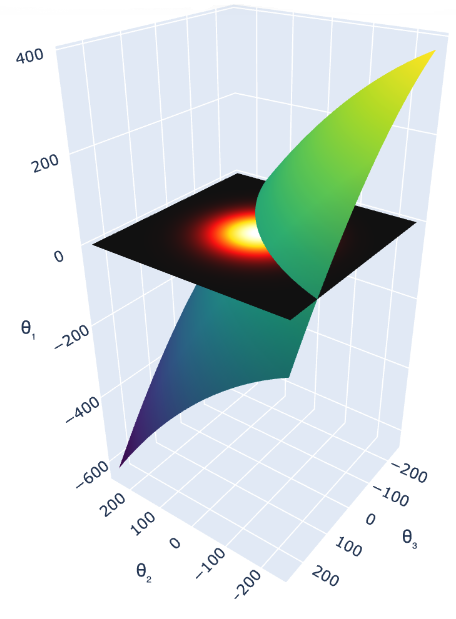}

}\subfloat[\foreignlanguage{british}{The eigenvalues used in determining the AS.\label{fig:The-eigenvalues-used-1}}]{\includegraphics[scale=0.4]{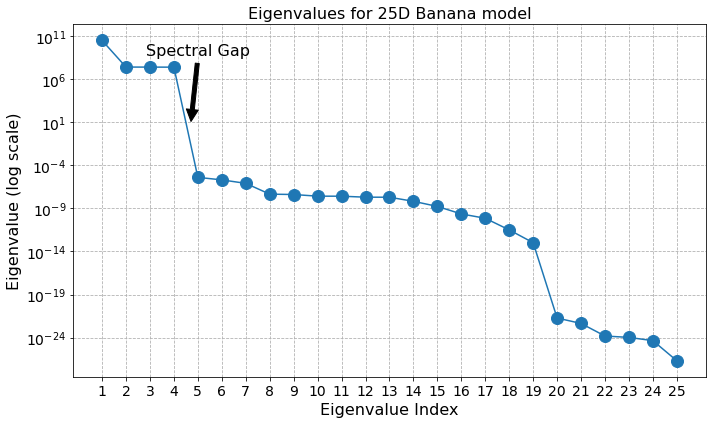}

}

\includegraphics[scale=0.5]{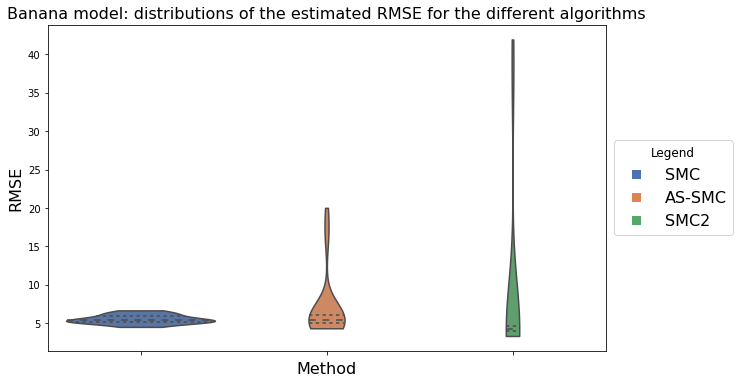}

\caption{\foreignlanguage{british}{Estimated RMSE of the posterior expectations over 50 runs.\label{fig:Estimated-RMSE-of-1}}}

\caption{\foreignlanguage{british}{The banana model: illustration and results.}}
\end{figure}

This model is very unrealistic. The artificial construction of a number
of variables that are not involved in the likelihood is tailor-made
for AS algorithms. However, this construction results in a ridge in
the posterior, which is possible in some real models. However, it
would be very unusual to see observe a ridge in the posterior that
is linear; for this reason we consider a variation on the model where
the ridge is not linear. Specifically, we consider instead the model
$y\sim\mathcal{N}\left(\sum_{i=1}^{d}\theta_{i}+b\sum_{j=1}^{k}\theta_{i}^{2},1\right)$,
for $1\leq k\leq d$. Figure \ref{fig:A-2d-slice-1} gives an illustration
of this model for $d=3$, $k=2$ and $b=0.001$. We call this model
the `banana' model, after the shape of the posterior it produces
in two dimensions. Whilst this model is again unrealistic, it provides
a useful test of the effectiveness of AS approaches outside of the
ideal case. The same models are studied in \citet{ripoli_particle_2024}.

\subsubsection{Results}

\begin{figure}
\subfloat[\foreignlanguage{british}{A 2d slice of the Gaussian prior on the horizontal plane $\theta_{1}=0$
together with the level surface of the likelihood in the particular
case $\theta_{1}+\theta_{2}+\theta_{3}=0$ (green plane).\label{fig:A-2d-slice}}]{\includegraphics[scale=0.4]{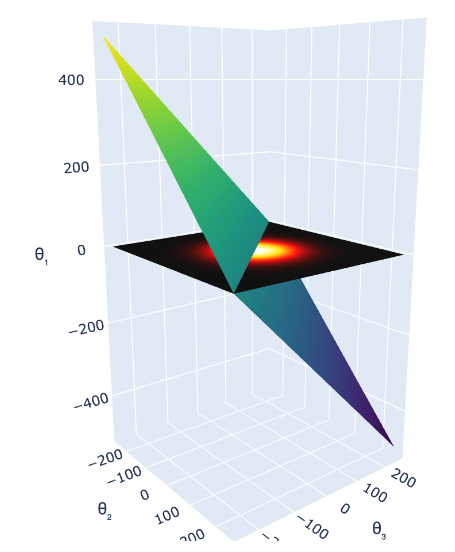}

}\subfloat[\foreignlanguage{british}{The eigenvalues used in determining the AS.\label{fig:The-eigenvalues-used}}]{\includegraphics[scale=0.4]{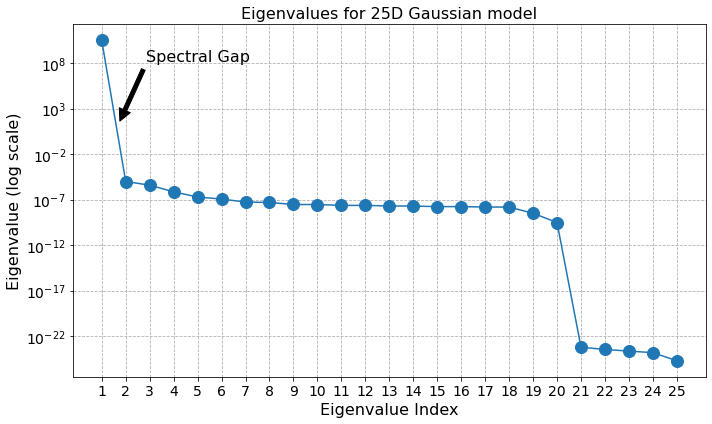}

}

\subfloat[\foreignlanguage{british}{Estimated RMSE of the posterior expectations over 50 runs.\label{fig:Estimated-RMSE-of}}]{\includegraphics[scale=0.5]{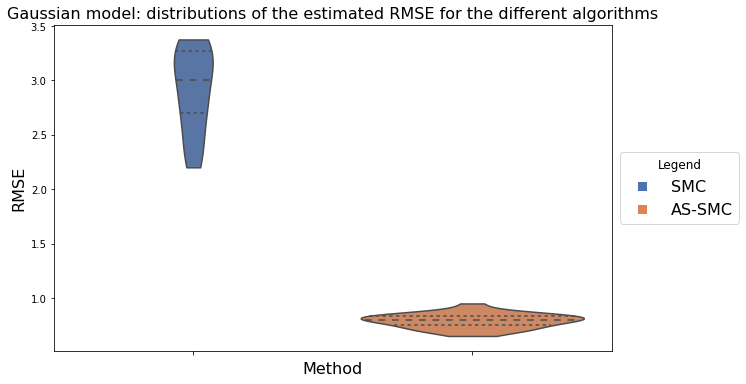}

}

\caption{\foreignlanguage{british}{The plane model: illustration and results.}}
\end{figure}

We take $d=25$ for the first model, and performed inference using
a standard SMC sampler and AS-SMC with a sequence of targets given
by annealing the likelihood. The eigenvalue approach led to the choice
of AS of dimension 1 (figure \ref{fig:The-eigenvalues-used}). We
used a fixed sequence of 25 targets for both algorithms, with the
sequence being determined by a preliminary run of an adaptive SMC
sampler. The standard SMC used $10^{4}$ particles; AS-SMC used the
same number of likelihood evaluations, taking $N_{a}=10^{3}$ and
$N_{i}=10$. Both algorithms used a Gaussian random walk proposal,
tuned using the adaptive approach from section \ref{subsec:MH-on}.
The proposal on inactive variables was taken to be the projected prior
$p_{i}$. We estimated the root mean squared error (RMSE) of the posterior
mean of each of the 25 parameters (approximating the truth as 0 for
each) from 50 runs of the algorithms. Figure \ref{fig:Estimated-RMSE-of}
compares these estimated RMSEs, where the distribution is across the
different parameters. We see that AS-SMC outperforms standard SMC
in this idealised case.

We now compared standard SMC, AS-SMC and AS-SMC$^{2}$ on the banana
model, for $d=25$, $k=3$ and $b=0.001$. The eigenvalue approach
led to the choice of AS of dimension 4 (figure \ref{fig:The-eigenvalues-used}).
This indicates a significant flaw in the use of active subspaces in
this example: the small curve in the `plane' introduced by the coefficient
$b=0.001$ results in four variables being grouped together in the
AS: it is only the variables that are absent from the likelihood that
are treated as inactive.

We again used a fixed sequence of 25 targets determined by a preliminary
run of an adaptive SMC. The standard SMC used $10^{4}$ particles
and AS-SMC and AS-SMC$^{2}$ used $N_{a}=10^{3}$ and $N_{i}=10$.
This gives AS-SMC the same number of likelihood evaluations as the
standard SMC, but due to the costlier MCMC moves the computational
cost of AS-SMC$^{2}$ is higher with $3\times10^{5}$ likelihood evaluations.
All algorithms used a Gaussian random walk proposal, tuned using the
adaptive approach from section \ref{subsec:MH-on}, and the prior
was always used as the proposal for the inactive variables. Figure
\ref{fig:Estimated-RMSE-of} compares the estimated RMSE of the estimated
posterior expectation over all parameters from 50 runs of the three
algorithms. In this example we observe poor performance from both
the AS-SMC and AS-SMC$^{2}$ algorithms. In both cases this is due
to the smaller number of (active space) particles used for these algorithms.
We use this smaller number of particles due to the additional computational
effort expended in the IS/SMC on the inactive space. This highlights
a weakness of AS approaches that build on the method of \citet{constantine_accelerating_2016}:
that we spend additional computational effort on simulating points
for the inactive variables, which are the variables that should be
easiest to simulate.

\selectlanguage{english}%

\selectlanguage{british}%

\subsection{Example of adaptive AS-SMC}

\begin{figure}
\includegraphics[scale=0.7]{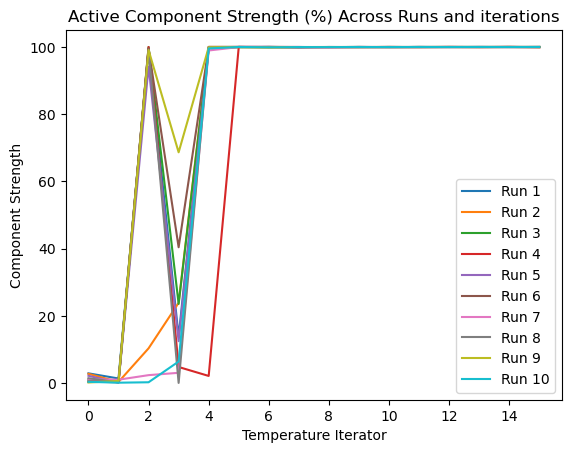}

\caption{The evolution of the estimated AS across the iterations of the adaptive
SMC.\label{fig:The-evolution-of}}

\end{figure}

To illustrate adaptive AS-SMC, we applied it to the toy example from
section \ref{subsec:Prior-and-posterior}. Our implementation used
an annealed sequence of distributions determined by a pilot run of
an adaptive SMC sampler. We performed 10 runs with $N_{a}=10^{3}$
active particles and $N_{i}=10$ inactive IS points, with a Gaussian
random walk proposal on the active variables whose variance was adapted
using the method in section \ref{subsec:MH-on}. Figure \ref{fig:The-evolution-of}
illustrates how the AS changes across the iterations of the adaptive
AS-SMC algorithm through showing, in the PCA on equation (\ref{eq:est_as}),
the proportion of variance explained by the active direction in the
posterior AS. However, although we see the expected evolution of the
AS in this example, we find that the adaptive AS-SMC algorithm has
poor performance compared to standard SMC. This is since, as can be
seen readily from the construction in section \ref{subsec:Prior-and-posterior},
there is no evidence that there are any inactive directions in this
example.

\section{Discussion\label{sec:Conclusions}}

In this paper we have introduced three SMC samplers that make use
of an AS. We have seen how the first of these, AS-SMC, can exhibit
improved performance over a standard SMC sampler in a model with variables
that are unidentifiable from the likelihood. The second method, adaptive
AS-SMC, can be used to estimate an appropriate AS at each iteration,
moving from the prior to the posterior. The third, AS-SMC$^{2}$,
addresses a different weakness of AS-SMC, substituting the IS step
in AS-SMC for an `internal' SMC move on the inactive variables in
order to avoid the exponential scaling of the variance of IS as the
dimension of the inactive variables increases.

SMC approaches can offer advantages over MCMC, since they use a population
of particles and are easier to tune. In this sense, AS-SMC has some
advantages over the current state-of-the-art AS approach, AS-MH. Adaptive
AS-SMC might improve on AS-SMC in the case where the posterior AS
is far from the prior AS, and where an `ideal' AS exists. \citet{parente_active_2020,zahm_certified_2022}
study such models: these papers also consider adaptive AS approaches,
although without theoretical guarantees. AS-SMC$^{2}$ has the potential
to provide lower variance estimates than AS-SMC, although it has a
higher computational cost.

All approaches considered in this paper are limited by the assumptions
underpinning active subspaces: that there is a subspace under a linear
transformation for which the likelihood is not informative. Outside
of this setting, it is not clear that AS-MH, AS-SMC, adaptive AS-SMC
or AS-SMC$^{2}$ will outperform a standard approach. Despite some
evidence in previous studies (e.g. \citet{constantine_accelerating_2016,parente_active_2020,zahm_certified_2022})
that using an AS can result in more efficient inference we are skeptical
that approaches that build on the marginal likelihood estimation approach
of \citet{constantine_accelerating_2016} will find wide applicability,
due to the poor performance of the methods when the linear transformation
is not suitable. This is studied further, in the context of MCMC,
in \citet{ripoli_particle_2024}.

\section*{Acknowledgements}

Leonardo Ripoli's work was supported by EPSRC grant EP/L016613/1 (the Centre for Doctoral Training in the Mathematics of Planet Earth). Richard G. Everitt was supported by EPSRC grant EP/W006790/1 and NERC grant NE/T00973X/1.

\appendix

\section{Full algorithms\label{sec:Full-algorithms}}

This section contains step-by-step descriptions of the algorithms
devised in the paper. Algorithm \ref{alg:adaptive_as_smc} contains the adaptive AS-SMC algorithm. Algorithm \ref{alg:inactive_smc} contains the internal SMC algorithm used on inactive variables, which is called by algorithm \ref{alg:as_smc2}.

\begin{algorithm}
\caption{Adaptive AS-SMC}\label{alg:adaptive_as_smc}
\tiny

Simulate $N_{a}$ points, $\left\{ \theta_{0}^{m} \right\}_{m=1}^{N_{a}} \sim p$ and set each weight $\omega_0^m = 1/N_{a}$;\

Find AS using points $\left\{ \theta_{0}^{m} \right\}_{m=1}^{N_{a}}$ in equation (\ref{eq:est_as}), yielding $A_0$ and $I_0$;\

\For{$m=1:N_{a}$}
{
	Set $a^m_{0}=A_{0}^{T}\left(\theta^m_{0}\right)$, $i^{1,m}_{0}=I_{0}^{T}\left(\theta^m_{0}\right)$ and $u_0^m = 1$;\

	\For{$n=2:N_{i}$}
	{         
		$i_{0}^{n,m} \sim q_{0,i}\left( \cdot \mid a_{0}^m \right) := p_{0,i}\left( \cdot \mid a_{0}^m \right)$;\
	}

	Set $w_0^{n,m} = 1/N_{i}$ for $n=1:N_i$;\
}

\For {$t=1:T$}
{
	Find AS using points $\left\{ a_{t-1}^{m} \left\{ i_{t-1}^{n,m} \right\}_{n=1}^{N_{i}} \right\}_{m=1}^{N_{a}}$ in equation (\ref{eq:smc_as}), yielding $A_t$ and $I_t$;\

	\For{$m=1:N_{a}$}
	{
		Set $a^m_{t-1 \rightarrow t}=A_t^{T}\left(A_{t-1} a^m_{t-1} + I_{t-1} i_{t-1}^{u_{t-1},m}\right)$, $i^{u_{t-1}^m,m}_{t-1 \rightarrow t}=I_t^{T}\left(A_{t-1}a^m_{t-1}+I_{t-1}i_{t-1}^{u^m_{t-1},m}\right)$ and $u_{t-1 \rightarrow t}^m = u_{t-1}^m$;\

		\For{$n=1:N_{i}$, $n \neq u_{t-1}^m$}
		{         
			$i_{t-1 \rightarrow t}^{n,m} \sim \kappa_{t-1,i}\left( \cdot \mid a_{t-1 \rightarrow t}^m \right)$;\
		}
	}

	\For(\tcp*[h]{reweight}){$m=1:N_{a}$}
	{
			\For{$n=1:N_{i}$}
			{         
				$\tilde{w}_{t}^{n,m}\left(a_{t-1 \rightarrow t}^{m},i_{t-1 \rightarrow t}^{n,m}\right)=\frac{p_{t,i}\left(i_{t-1 \rightarrow t}^{n,m}\mid a_{t-1 \rightarrow t}^{m}\right)l_{1:t}\left(A_t a_{t-1 \rightarrow t}^{m}+I_t i_{t-1 \rightarrow t}^{n}\right)}{q_{t,i}\left(i_{t-1 \rightarrow t}^{n,m}\mid a_{t-1 \rightarrow t}^{m}\right)}$;\

$\tilde{w}_{t-1}^{n,m}\left(a_{t-1 \rightarrow t}^{m},i_{t-1 \rightarrow t}^{n,m}\right)=\frac{p_{t,i}\left(i_{t-1 \rightarrow t}^{n,m}\mid a_{t-1 \rightarrow t}^{m}\right)l_{1:t-1}\left(A_t a_{t-1 \rightarrow t}^{m} + I_t i_{t-1 \rightarrow t}^{n}\right)}{\kappa_{t-1,i}\left(i_{t-1 \rightarrow t}^{n,m}\mid a_{t-1 \rightarrow t}^{m}\right)}$;\
			}
			
			\[
			\tilde{\omega}_{t}^{m} = \omega_{t-1}^{m}\frac{\prod_{j=1}^{N_{i}}q_{t,i}\left(\theta_{i_{t}}^{j}\mid\theta_{a_{t}}\right)\sum_{n=1}^{N_{i}}\tilde{w}_{t}^{n,m}\left(a_{t-1\rightarrow t}^{m},i_{t-1\rightarrow t}^{n,m}\right)}{\prod_{j=1}^{N_{i}}\kappa_{t-1,i}\left(\theta_{i_{t}}^{j}\mid\theta_{a_{t}}\right)\sum_{n=1}^{N_{i}}\tilde{w}_{t-1}^{n,m}\left(a_{t-1\rightarrow t}^{m},i_{t-1\rightarrow t}^{n,m}\right)};
			\]
	}

	$\left\{ \omega^m_{t} \right\}_{m=1}^{N_{a}} \leftarrow \mbox{ normalise}\left( \left\{ \tilde{\omega}^m_{t} \right\}_{m=1}^{N_{a}} \right);$

	\For{$m=1:N_{a}$}
	{
		$\left\{ w^{n,m}_{t} \right\}_{n=1}^{N_{i}} \leftarrow \mbox{ normalise}\left( \left\{ \tilde{w}^{n,m}_{t} \right\}_{n=1}^{N_{i}} \right);$
	}

	\For{$m=1:N_{a}$}
	{
		$u_t^{m} \sim\mathcal{M}\left( \left( w_t^{1,m}, ..., w_t^{N_i,m} \right) \right)$;
	}

	\If(\tcp*[h]{resample and move}) {some degeneracy condition is met}
	{
		\For{$m=1:N_{a}$}
		{
			Simulate $\left(\theta_{t,a_t}^{m}, \theta_{t,i_t}^{1:N_{i},m} \right)$ from the mixture distribution
			\[
			\sum_{j=1}^{N_{a}} \omega_t^{j} K_{t,a}\left\{ \cdot \mid \left(a_{t-1\rightarrow t}^{j},i_{t-1\rightarrow t}^{1:N_{i},j}\right) \right\},
			\]
			where $K_{t,a}$ is an AS-MH move, i.e.:

			$j^* \sim \mathcal{M}\left( \left\{ \omega_t^{j} \right\}_{j=1}^{N_{a}} \right)$;\

			$a_{t}^{*m} \sim q_{t,a}\left(\cdot\mid a_{t-1 \rightarrow t}^{j^*} \right)$;\

			\For {$n=1:N_i$}
			{
		
				$i_{t}^{*n,m} \sim q_{t,i} \left(\cdot \mid a_{t}^{*m} \right);$\
		
				\[
				\tilde{w}_t^{*n,m} \left(a_{t}^{*m},i_{t}^{*n,m}\right) = \frac{p_{t,i}\left(i^{*n,m}_{t} \mid a^{*m}_{t}\right) l_{1:t}\left( A_t a_{t}^{*m} + I_t i_{t}^{*n,m}\right)}{q_{t,i}\left( i_{t}^{*n,m} \mid a_{t}^{*m} \right)};
				\]\
			}
		
			$u_t^{*m} \sim\mathcal{M}\left( \left( w_t^{*1,m}, ..., w_t^{*N_i,m} \right) \right),$ where for $n=1:N_i$
			\[
			w_t^{*n,m} = \frac{\tilde{w}_t^{*n,m}}{\sum_{p=1}^{N_{i}} \tilde{w}_t^{*p,m}};\
			\]
		
		   Set $\left(a_{t}^m, \left\{ i_{t}^{n,m}, \tilde{w}_t^{n,m} \right\}_{n=1}^{N_i}, u_t^m \right) = \left(a_{t}^{*m},\left\{ i_{t}^{*n,m}, \tilde{w}_t^{*n,m} \right\}_{n=1}^{N_i}, u_t^{*m} \right)$ with probability
			\[
			\alpha_{t,a_t}^m = 1\wedge\frac{p_{t,a}\left(a^{*m}_{t}\right) \sum_{n=1}^{N_{i}}\tilde{w}_{t}^{*n,m}\left(a_{t}^{*m},i_{t}^{*n,m}\right)}{p_{t,a}\left(a^{j^*}_{t-1 \rightarrow t}\right) \sum_{n=1}^{N_{i}}\tilde{w}_{t}^{n,j^*}\left(a_{t-1 \rightarrow t}^{j^*},i_{t-1 \rightarrow t}^{n,j^*}\right)}\frac{q_{t,a}\left(a_{t}^{j^*} \mid a_{t}^{*m}\right)}{q_{t,a}\left(a_{t,a}^{*m}\mid a_{t}^{j^*} \right)};
			\]\
			
			Else let $\left(a_{t}^m, \left\{ i_{t}^{n,m}, \tilde{w}_t^{n,m} \right\}_{n=1}^{N_i}, u_t^m \right) = \left(a_{t}^{j^*},\left\{ i_{t}^{n,j^*}, \tilde{w}_t^{n,j^*} \right\}_{n=1}^{N_i}, u_t^{j^*}\right)$;\
		}

		$\omega^m_{t} = 1/N_{a}$ for $m=1:N_{a}$;
	}

}
\end{algorithm}

\begin{algorithm} \caption{SMC on inactive variables for a given $a$ and $t$.}\label{alg:inactive_smc}

Simulate $N_{i}$ points, $\left\{ i_{0}^{n} \right\}_{n=1}^{N_{i}} \sim p_i\left( \cdot \mid a \right)$ and set each weight $w_0^n = 1/N_i$;\

\For {$s=1:t$}
{
	\For(\tcp*[h]{reweight}) {$n=1:N_i$}
	{
		\eIf {$s=1$}
		{
			\[
			\tilde{w}^n_{s} = w^n_{s-1} l_{1:s}\left(B_{a} a+B_{i} i^{n}\right);
			\]
		}
		{
			\[
			\tilde{w}^n_{s} = w^n_{s-1} \frac{l_{1:s}\left(B_{a} a+B_{i} i_{s-1}^{n}\right)}{l_{1:s-1}\left(B_{a}a+B_{i}i_{s-1}^{n}\right)};
			\]
		}
	}

	$\left\{ w^n_{s} \right\}_{n=1}^{N_{i}} \leftarrow \mbox{ normalise}\left( \left\{ \tilde{w}^n_{s} \right\}_{n=1}^{N_{i}} \right)$;

	If $s=t$, go to line 32;

	\For{$n=1:N_i$}
	{
		Simulate the index $v^n_{s-1} \sim \mathcal{M}\left( \left( w_s^1, ..., w_s^{N_i} \right) \right)$ of the ancestor of particle $n$;
	}

	\If(\tcp*[h]{resample}) {some degeneracy condition is met}
	{
		\For{$n=1:N_i$}
		{
			Set $i^{n}_{s} = i^{v^n_{s-1}}_{s-1}$;
		}
		$w^n_{s} = 1/N_i$ for $n=1:N_i$;
	}
	\Else
	{
		\For{$n=1:N_i$}
		{
			Set $i^{n}_{s} = i^{n}_{s-1}$;
		}
	}
	
	\For(\tcp*[h]{move}) {$n=1:N_i$}
	{
		Simulate $i^{n*}_{s} \sim q_{t,i} \left( \cdot \mid i^{n}_{s}, a \right)$;
		
		Set $i^{n}_{s} = i^{n*}_{s}$ with probability

		\[
		1\wedge\frac{l_{1:s}\left(B_{a}a+B_{i}i^{n*}_{s}\right)p_i\left( i^{n*}_{s} \mid a \right) q_{t,i} \left( i^{n}_{s} \mid a \right)}{l_{1:s}\left(B_{a}a+B_{i}i^{n}_{s}\right)p_i\left( i^{n}_{s} \mid a \right) q_{t,i} \left( i^{n*}_{s} \mid a \right)},
		\]
	}
}
Estimate $l_{t,a}\left( a \right)$ using
\[
\overline{l}_{t,a}\left(a\right) = \prod_{s=1}^t \sum_{n=1}^{N_{i}} \tilde{w}^n_{s}.
\]
\end{algorithm}

\begin{algorithm}
\caption{Active subspace SMC$^2$}\label{alg:as_smc2}
\small

Simulate $N_{a}$ points, $\left\{ \theta_{0}^{m} \right\}_{m=1}^{N_{a}} \sim p$ and set each weight $\omega_0^m = 1/N_{a}$;\

Find AS using points $\left\{ \theta_{0}^{m} \right\}_{m=1}^{N_{a}}$ in equation (\ref{eq:est_as}), yielding $B_a$ and $B_i$;\

\For{$m=1:N_{a}$}
{
	Set $a^m_{0}=B_{a}^{T}\left(\theta^m_{0}\right)$, $i^{1,m}_{0}=B_{i}^{T}\left(\theta^m_{0}\right)$ and $u_0^m = 1$;\

	\For{$n=2:N_{i}$}
	{         
		$i_{0}^{n,m} \sim q_{0,i}\left( \cdot \mid a_{0}^m \right) := p_i\left( \cdot \mid a_{0}^m \right)$;\
	}

	Set $w_0^{n,m} = 1/N_{i}$ for $n=1:N_i$;\
}

\For {$t=1:T$}
{
	\For(\tcp*[h]{reweight}){$m=1:N_{a}$}
	{
			\eIf {$t=1$}
			{
				Simulate $\left(i_{t}^{1: N_{i}, m}, v_{t-1}^{1: N_{i}, m}\right)$ using lines 3-30 (ignoring line 11) of algorithm \ref{alg:inactive_smc} taking $s$ in these lines equal to $t$, then compute
				\[
				\widehat{l_{t,a}\left(a_{t-1}^{m}\right)}= \sum_{n=1}^{N_{i}} \tilde{w}^{n,m}_{t};
				\]
				\[
				\tilde{\omega}_t^{m} = \omega_{t-1}^{m} \widehat{l_{t,a}\left(a_{t-1}^{m}\right)};
				\]
			}
			{
				Simulate $\left(i_{t}^{1: N_{i}, m}, v_{t-1}^{1: N_{i}, m}\right)$ using lines 3-30 (ignoring line 11) of algorithm \ref{alg:inactive_smc} taking $s$ in these lines equal to $t$, then compute:
				\[
				\widehat{\frac{l_{t,a}\left({a_{t-1}^{m}}\right)}{l_{t-1,a}\left(a_{t-1}^{m}\right)}}= \sum_{n=1}^{N_{i}} \tilde{w}^{n,m}_{t};
				\]
				\[
				\tilde{\omega}_t^{m} = \omega_{t-1}^{m} \widehat{\frac{l_{t,a}\left({a_{t-1}^{m}}\right)}{l_{t-1,a}\left(a_{t-1}^{m}\right)}};
				\]
			}
	}

	$\left\{ \omega^m_{t} \right\}_{m=1}^{N_{a}} \leftarrow \mbox{ normalise}\left( \left\{ \tilde{\omega}^m_{t} \right\}_{m=1}^{N_{a}} \right);$

	\If(\tcp*[h]{resample and move}) {some degeneracy condition is met}
	{
		\For{$m=1:N_{a}$}
		{
			Simulate $\left(a_{t}^{m}, i_{1:t}^{1:N_{i},m}, v_{1:t-1}^{1:N_{i},m} \right)$ from the mixture distribution
			\[
			\sum_{j=1}^{N_{a}} \omega_t^{j} K_{t,a}\left\{ \cdot \mid \left(a_{t-1}^{j},i_{t}^{1:N_{u},j},v_{t-1}^{1:N_{i},j}\right) \right\},
			\]
			where $K_{t,a}$ is an AS-PMMH move, i.e.:

			$j^* \sim \mathcal{M}\left( \left\{ \omega_t^{j} \right\}_{j=1}^{N_{\theta}} \right)$, then $a^* \sim q_{t,a} \left( \cdot \mid a_{t-1}^{j^*} \right)$, then run algorithm \ref{alg:inactive_smc} up to target $t$ conditional on $a^*$.

			Set $a_{t}^{m} = a^*$ and $i^{n,m}_{1:t}, v^{n,m}_{1:t-1}$ and $\tilde{w}^{n,m}_{1:t}$ to be the variables and unnormalised weights generated when running algorithm \ref{alg:inactive_smc} with probability 

			\[
			1\wedge\frac{p_a\left(a^{*}\right)}{p_a\left(a_{t-1}^{j^*}\right)}\frac{q_{t,a}\left(a_{t-1}^{j^*} \mid a^{*}\right)}{q_{t,a}\left(a^{*}\mid a_{t-1}^{j^*} \right)} \frac{\overline{l}_{t,a}\left(a^*\right)}{\prod_{t=1}^T \sum_{n=1}^{N_{i}} \tilde{w}^{n,j^*}_{t}},
			\]
			where $\overline{l}_{t,a}\left(a^*\right)$ given by algorithm \ref{alg:inactive_smc} run conditional on $\theta_a^*$;

			Else set $a_{t}^{m} = a_{t-1}^{j^*}$, $\tilde{w}^{n,m}_{1:t} = \tilde{w}^{n,j^*}_{1:t}$, $i^{n,m}_{1:t}=i^{n,j^*}_{1:t}$ and $v^{n,m}_{1:t-1}=v^{n,j^*}_{1:t-1}$.
		}

		$\omega^m_{t} = 1/N_{a}$ for $m=1:N_{a}$;
	}

}
\end{algorithm}

\section{Special cases for adaptive AS-SMC\label{sec:Special-cases-for}}

This section describes the special cases of adaptive AS-SMC when no
inactive variables are found in either iteration $t-1$ or iteration
$t$.

\subsection{Moving to no inactive variables\label{subsec:Moving-to-no}}

When there are no inactive variables at iteration $t$, $a=\theta$,
and the target distribution is chosen to be
\[
\pi_{t}\left(\theta\right)=\frac{1}{Z_{t}}p\left(\theta\right)l_{1:t}\left(\theta\right).
\]
We first consider the reprojection step at the beginning of the SMC
iteration, when moving from inactive variables existing at iteration
$t-1$, but not existing at iteration $t$. This step is more straightforward
than previously. The desired target distribution for the move is
\[
\pi_{t-1}^{*}\left(u,\theta\right)=\frac{1}{N_{i}}\frac{1}{Z_{t-1}}p\left(\theta\right)l_{1:t-1}\left(\theta\right),
\]
and the extended target then
\[
\pi_{t-1}^{*}\left(u,\theta\right)L\left(\left\{ i_{t-1}^{n}\right\} _{n=1,n\neq u}^{N_{i}}\mid u,\theta\right).
\]
The proposal in the reprojection is $\pi_{t-1}^{*}\left(u,a_{t-1},\left\{ i_{t-1}^{n}\right\} _{n=1}^{N_{i}}\right)$,
i.e. the target distribution from the previous step. The proposal
takes a point $\left(u_{t-1},a_{t-1},\left\{ i_{t-1}^{n}\right\} _{n=1}^{N_{i}}\right)$
from this target and transforms it using
\begin{equation}
\theta_{t-1\rightarrow t}=G_{t-1\rightarrow t,a}\left(u_{t-1},a_{t-1},\left\{ i_{t-1}^{n}\right\} _{n=1}^{N_{i}}\right)=A_{t-1}a_{t-1}+I_{t-1}i_{t-1}^{u_{t-1}}\label{eq:transform_to_no_inactive}
\end{equation}
\[
u_{t-1\rightarrow t}=u_{t-1}.
\]
The $L$ kernel is again chosen so that the importance weights are
1, and the indexing variable is again discarded. The result is that
this step simplifies to applying the transform in equation (\ref{eq:transform_to_no_inactive})
to each particle.

The weight update then follows as
\begin{eqnarray*}
\tilde{\omega}_{t}^{m} & = & \frac{p\left(\theta_{t-1\rightarrow t}^{m}\right)l_{1:t}\left(\theta_{t-1\rightarrow t}^{m}\right)}{p\left(\theta_{t-1\rightarrow t}^{m}\right)l_{1:t-1}\left(\theta_{t-1\rightarrow t}^{m}\right)}\\
 & = & l_{t}\left(\theta_{t-1\rightarrow t}^{m}\right),
\end{eqnarray*}
 and the MCMC move at iteration $t$ is then a standard MH step.

\subsection{Moving from no inactive variables}

When there are no inactive variables at iteration $t-1$ and there
are again no inactive variables at iteration $t$, we simply follow
section \ref{eq:transform_to_no_inactive}, but do not need to the
transform in equation (\ref{eq:transform_to_no_inactive}), i.e. the
algorithm is a standard SMC sampler. When there no inactive variables
at iteration $t-1$, but there are inactive variables at iteration
$t$ we mostly follow section \ref{subsec:Construction}, except that
the conditional IS step at the beginning of the SMC iteration is slightly
simpler.

As the target distribution for this IS step, we may use $\pi_{t-1}\left(a_{t},\left\{ i_{t}^{n}\right\} _{n=1}^{N_{i}}\right)$:
the indexing variable $u$ is not required. We use the transformation
\[
a_{t-1\rightarrow t}:=G_{t-1\rightarrow t,a}\left(\theta_{t-1}\right):=A_{t}^{T}\left(\theta_{t-1}\right)
\]
\[
i_{t-1\rightarrow t}:=G_{t-1\rightarrow t,i}\left(\theta_{t-1}\right):=I_{t}^{T}\left(\theta_{t-1}\right)
\]
on the point $\theta_{t-1}$, then, if $N_{i}>1$, need to generate
the remaining variables $i_{t-1\rightarrow t}^{n}$ for $n=2:N_{i}$.
We simulate each from $q_{t-1\rightarrow t}$ and see straightforwardly
that this results in an importance weight of 1.

\section{Justification for AS-SMC$^{2}$\label{sec:Justification-for-AS-SMC}}

This section follows exactly the corresponding argument in \citet{Chopin2013}:
here we sketch the main ideas, with the full argument being found
in that paper. 

At iteration $t$ we require our algorithm to have as one of its marginals
the target distribution
\[
\pi_{t}\left(a,i\right)=\frac{p_{a}\left(a\right)p_{i}\left(i\mid a\right)l_{1:t}\left(B_{a}a+B_{i}i\right)}{Z_{t}},
\]
with marginal distribution

\[
\pi_{t,a}\left(a\right)=\frac{p\left(a\right)l_{t,a}\left(a\right)}{Z_{t}},
\]
where
\[
l_{t,a}\left(a\right)=\int_{i}p_{i}\left(i\mid a\right)l_{1:t}\left(B_{a}a+B_{i}i\right)di.
\]

At $t=0$, $\ensuremath{\ensuremath{\left\{ a_{0}^{m}\right\} _{m=1}^{N_{a}}\sim p_{a}}}$
and $\left\{ i_{0}^{n,m}\right\} _{n=1}^{N_{i}}\sim p_{i}\left(\cdot\mid a_{0}^{m}\right)$
for $m=1:N_{u}$. The target at iteration 0 is
\[
p\left(a\right)\prod_{n=1}^{N_{i}}p_{i}\left(i_{0}^{n}\mid a\right),
\]
from which we simulate $N_{a}$ $a$-points and $N_{i}$ $i$-points
for every $a$. At $t=1$ each particle is assigned the weight

\begin{equation}
\hat{l}_{1,a}\left(a\right)=\frac{1}{N_{i}}\sum_{n=1}^{N_{i}}l_{1}\left(B_{a}a+B_{i}i_{0}^{n}\right).\label{eq:est}
\end{equation}
We introduce notation for the distribution of the $i$-variables generated
at iteration 0 used to estimate $l_{1,a}$: $\psi_{0}\left(\left\{ i_{0}^{n}\right\} _{n=1}^{N_{i}}\mid a\right)=\prod_{n=1}^{N_{i}}p_{i}\left(i_{0}^{n}\mid a\right)$.
The target distribution at $t=1$ being
\begin{equation}
\pi_{1}\left(a,\left\{ i_{0}^{n}\right\} _{n=1}^{N_{i}}\right)=p_{a}\left(a\right)\psi_{0}\left(\left\{ i_{0}^{n}\right\} _{n=1}^{N_{i}}\mid a\right)\frac{\hat{l}_{1,a}\left(a\right)}{Z_{1}}.\label{eq:target1}
\end{equation}
This results in the weight update in equation (\ref{eq:est}). We
can rewrite the target as
\begin{eqnarray}
\pi_{1}\left(a,\left\{ i_{0}^{n}\right\} _{n=1}^{N_{i}}\right) & = & \frac{p_{a}\left(a\right)}{Z_{1}}\prod_{n=1}^{N_{i}}p_{i}\left(i_{0}^{n}\mid a\right)\left(\frac{1}{N_{i}}\sum_{n=1}^{N_{i}}l_{1}\left(B_{a}a+B_{i}i_{0}^{n}\right)\right)\nonumber \\
 & = & \frac{1}{N_{i}}\sum_{n=1}^{N_{i}}\frac{p_{a}\left(a\right)}{Z_{1}}p_{i}\left(i_{0}^{n}\mid a\right)l_{1}\left(B_{a}a+B_{i}i_{0}^{n}\right)\left(\prod_{\substack{j=1\\
j\neq n
}
}^{N_{i}}p_{i}\left(i_{0}^{j}\mid a\right)\right)\nonumber \\
 & = & \frac{\pi_{1,a}\left(a\right)}{N_{i}}\sum_{n=1}^{N_{i}}\pi_{t,i}\left(i_{0}^{n}\mid a\right)\left(\prod_{\substack{j=1\\
j\neq n
}
}^{N_{i}}p_{i}\left(i_{0}^{j}\mid a\right)\right)\label{eq:target1_alt}
\end{eqnarray}
where we used the result
\[
p_{a}\left(a\right)p_{i}\left(i_{0}^{n}\mid a\right)l_{1}\left(B_{a}a+B_{i}i_{0}^{n}\right)=Z_{1}\pi_{1,a}\left(a\right)\pi_{t,i}\left(i_{0}^{n}\mid a\right).
\]
In the marginals of equation (\ref{eq:target1_alt}) we have the target
at iteration 1 of $\pi_{1}\left(a,i\mid y\right)=\pi_{1,a}\left(a\right)\pi_{1,i}\left(i_{0}^{n}\mid a\right)$.

For $t\geq2$, similarly to equation \ref{eq:target1}, we again have
that our target distribution is defined to be the prior, multiplied
by the likelihood estimate, multipled by the distribution of the variables
used in the likelihood estimator, multiplied by the normalising constant
(which again follows from the unbiasedness of the likelihood estimator).

Let $\psi_{t-1}$ be the distribution of all of the random variables
generated by the internal SMC up to time $t$.
\begin{equation}
\pi_{t}\left(a,\left\{ i_{0:t-1}^{n},v_{0:t-1}^{n}\right\} _{n=1}^{N_{i}}\right)=p_{a}\left(a\right)\psi_{t-1}\left(\left\{ i_{0:t-1}^{n},v_{0:t-1}^{n}\right\} _{n=1}^{N_{i}}\mid a\right)\frac{\hat{l}_{t,a}\left(a\right)}{Z_{t}},\label{eq:targett}
\end{equation}
Similarly to the $t=1$ case, we may rearrange equation (\ref{eq:targett})
to see that $\pi_{t}\left(a,i\right)$ is included in its marginals:

\begin{align*}
\pi_{t}\left(a,\left\{ i_{0:t-1}^{n},v_{0:t-1}^{n}\right\} _{n=1}^{N_{i}}\right) & =\frac{\pi_{t,a}\left(a\right)}{N_{i}}\times\sum_{n=1}^{N_{i}}\frac{\pi_{t,i}\left(\mathbf{i}_{1:t-1}^{n}\mid a\right)}{N_{u}^{t-1}}\\
 & \left(\prod_{\substack{j=1\\
j\neq\mathbf{h}_{t}^{n}\left(0\right)
}
}^{N_{i}}p_{i}\left(i_{0}^{j}\mid a\right)\right)\left(\prod_{s=2}^{t}\prod_{\substack{j=1\\
j\neq\mathbf{h}_{t}^{n}\left(s-1\right)
}
}^{N_{i}}w_{s-1}^{v_{s-1}^{j}}K_{s-1,i}\left(i_{s-1}^{j}\mid i_{s-2}^{v_{s-2}^{j}},a\right)\right),
\end{align*}
where $\mathbf{i}_{1:t-1}^{n}$ and $\mathbf{h}_{t}^{n}$ are deterministic
functions of $\left\{ i_{0:t-1}^{n}\right\} _{n=1}^{N_{i}}$ and $\left\{ v_{0:t-1}^{n}\right\} _{n=1}^{N_{i}}$:
$\mathbf{h}_{t}^{n}=\left(\mathbf{h}_{t}^{n}\left(0\right),...,\mathbf{h}_{t}^{n}\left(t-1\right)\right)$
denotes the index history of $v_{t-1}^{n}$, i.e. $\mathbf{h}_{t}^{n}\left(t-1\right)=n$
and $\mathbf{h}_{t}^{n}\left(s\right)=v^{\mathbf{h}_{t}^{n}\left(s+1\right)}$,
recursively for $s=t-2,...,0$, and $\mathbf{i}_{1:t-1}^{n}=\left(\mathbf{i}_{1:t-1}^{n}(0),...,\mathbf{i}_{1:t-1}^{n}(t-1)\right)$
denotes the state trajectory of particle $i_{t-1}^{n}$, i.e. $\mathbf{i}_{1:t-1}^{n}(s)=i_{s}^{h_{t}^{n}(s)}$,
for $s=0,...,t-1$.

For the remainder of the proof we follow \citet{Chopin2013}, with
the one difference in the notation from that paper that here the index
of the $i$ variable is one fewer: i.e. equation (\ref{eq:targett})
uses $\psi_{t-1}\left(\left\{ i_{0:t-1}^{n},v_{0:t-1}^{n}\right\} _{n=1}^{N_{i}}\mid a\right)$,
whereas the equivalent in \citet{Chopin2013} would be $\psi_{t}\left(\left\{ i_{1:t}^{n},v_{1:t-1}^{n}\right\} _{n=1}^{N_{i}}\mid a\right)$.
The reason is that here the weight update in the internal SMC only
involves the values of the particles from the previous iteration.

\bibliographystyle{mychicago}
\bibliography{smc_active_subspace}

\end{document}